\ifx\onecol\undefined
\documentclass[journal, twocolumn]{IEEEtran}
\else
\documentclass[12pt,draftclsnofoot,journal,onecolumn]{IEEEtran}
\fi

\usepackage{amsfonts}
\usepackage{amsmath,amssymb}
\usepackage{acronym}  
\usepackage{algorithm}
\usepackage{algorithmic}
\usepackage{balance}
\usepackage{bm}
\usepackage{bbm}
\usepackage{booktabs}
\usepackage{color, soul}
\usepackage{cite}
\usepackage{graphicx}
\usepackage{indentfirst}
\usepackage{setspace}
\usepackage{tikz}
\usetikzlibrary{arrows}
\usepackage{subfigure}
\usepackage[amsmath,thmmarks]{ntheorem}
\usepackage{theorem}
\usepackage{hyperref}
\hypersetup{hidelinks, 
colorlinks=true,
allcolors=black,
pdfstartview=Fit,
breaklinks=true}

\theoremheaderfont{~~~\it}\theorembodyfont{\upshape}%
\theoremstyle{nonumberplain}
\theoremseparator{}
\theoremsymbol{\rule{1ex}{1ex}}

\acrodef{EAR}[EAR]{element activation ratio}
\acrodef{SNR}[SNR]{signal-to-noise ratio}
\acrodef{TC}[TC]{transmission coefficients}
\acrodef{US-RIS}[US-RIS]{user-side RIS}
\acrodef{BSS-RIS}[BSS-RIS]{base-station-side RIS}
\acrodef{DoF}[DoF]{degree of freedom}
\acrodef{FPGA}[FPGA]{field programmable gate array}
\acrodef{RF}[RF]{radio-frequency}
\acrodef{UE}[UE]{user equipment}
\acrodef{BS}[BS]{base station}
\acrodef{RIS}[RIS]{reconfigurable intelligent surface}
\acrodef{DL}[DL]{downlink}
\acrodef{TA}[TA]{transmit antenna}
\acrodef{RA}[RA]{receive antenna}
\acrodef{LoS}[LoS]{line-of-sight}
\acrodef{UL-TBF}[UL-TBF]{uplink transmit beamforming}
\acrodef{TPS}[TPS]{transmit phase shifter}
\acrodef{RC}[RC]{receiver combining}
\acrodef{AWGN}[AWGN]{additive white Gaussian noise}
\acrodef{DFT}[DFT]{discrete Fourier transform}
\acrodef{IRF}[IRF]{interference random field}
\acrodef{MISO}[MISO]{multiple-input single-output}
\acrodef{CSI}[CSI]{channel state information}
\acrodef{CRLB}[CRLB]{Cramer-Rao lower bound}
\acrodef{UPA}[UPA]{uniform planar array}

\def \H {^{\rm H}}

\def \T {^{\rm T}}

\def \Pmax {P_{\text{max}}}
\def \Pmin {p_{\text{min}}}

\def \diag {\text{diag}}

\def \arg {\text{arg}}
\def \j {\text{j}}

\def \tr {\text{tr}}
\def \SE {\text{SE}}
\def \EE {\text{EE}}

\newcommand{\RNum}[1]{\uppercase\expandafter{\romannumeral #1\relax}}
\newcommand{\myincludegraphics}[2][width=\linewidth]{\includegraphics[#1]{#2}}

\ifCLASSINFOpdf
\else
\fi
\begin{document}
\title{Enhancing Energy Efficiency for Reconfigurable Intelligent Surfaces with Practical Power Models}
\author{{Zhiyi Li, Jida Zhang, Jieao Zhu, Shi Jin, and Linglong Dai}
\thanks{This work was supported in part by the National Key Research and Development Program of China (Grant No.2020YFB1807201), in part by the National Natural Science Foundation of China (Grant No. 62031019), and in part by the European Commission through the H2020-MSCA-ITN META WIRELESS Research Project (Grant No. 956256). An earlier version of this paper was presented in part at the IEEE International Wireless Communications and Mobile Computing Conference (IEEE IWCMC'23), Marrakech, Morocco, June 2023~\cite{li2023ris}. \it{(Corresponding author: Linglong Dai.)}}
\thanks{Z. Li, J. Zhang, J. Zhu, and L. Dai are with the Department of Electronic Engineering, Tsinghua University, Beijing 100084, China as well as the Beijing National Research Center for Information Science and Technology, Beijing, China (BNRist) (e-mails: \{lizhiyi20, zhang-jd20, zja21\}@mails.tsinghua.edu.cn, daill@tsinghua.edu.cn).}
\thanks{S. Jin is with the National Mobile Communications Research Laboratory, Southeast University, Nanjing 210096, China (e-mail: jinshi@seu.edu.cn).}
}

\maketitle

\begin{abstract}
    Reconfigurable intelligent surfaces (RISs) are widely considered a promising technology for future wireless communication systems.
    As an important indicator of RIS-assisted communication systems in green wireless communications, energy efficiency (EE) has recently received intensive research interest as an optimization target. 
    However, most previous works have ignored the different power consumption between ON and OFF states of the PIN diodes attached to each RIS element. 
    This oversight results in extensive unnecessary power consumption and reduction of actual EE due to the inaccurate power model. 
    To address this issue, in this paper, we first utilize a practical power model for a RIS-assisted multi-user multiple-input single-output (MU-MISO) communication system, which takes into account the difference in power dissipation caused by ON-OFF states of RIS's PIN diodes.   
    Based on this model, we formulate a more accurate EE optimization problem. 
    However, this problem is non-convex and has mixed-integer properties, which poses a challenge for optimization. 
    To solve the problem, an effective alternating optimization (AO) algorithm framework is utilized to optimize the base station and RIS beamforming precoder separately. 
    To obtain the essential RIS beamforming precoder, we develop two effective methods based on maximum gradient search and SDP relaxation respectively. 
    Theoretical analysis shows the exponential complexity of the original problem has been reduced to polynomial complexity. 
    Simulation results demonstrate that the proposed algorithm outperforms the existing ones, leading to a significant increase in EE across a diverse set of scenarios.

\end{abstract}

\begin{IEEEkeywords}
Reconfigurable intelligent surface (RIS), energy efficiency (EE), non-convex mixed-integer programming, semi-definite programming (SDP). 
\end{IEEEkeywords}

\section{Introduction}
\label{Introduction}

Recently, a new concept called \ac{RIS} has attracted enormous attention and academic interest in wireless communications society.
Specifically, \ac{RIS} is a large reflection array composed of numerous nearly passive elements. By controllably tuning the phase-shifts of the incident signals, these elements are capable of cooperatively reflecting the signals towards desired directions with high beamforming gain~\cite{Di2020smart, Pan2021reconfigurable, Wu2020towards}. 
Due to its unique characteristics, \ac{RIS} is expected to provide various performance improvements in wireless communications, including overcoming blockages, enhancing spectrum efficiency, and reducing energy consumption~\cite{Liu2021compact, Basar2019wireless, Wu2019intelligent}. 
Among these, the reduction of energy consumption in \ac{RIS}-assisted communication systems has been gaining increasing research interest, especially considering the growing demands for low-power massive connections and green radio in future wireless communications~\cite{wu2017overview}.

To realize energy-efficient \ac{RIS}-aided communications, it is of practical importance to study the power consumption of RIS's hardware. 
Usually, \acp{RIS} are manufactured with massive number of nearly passive elements, such as PIN diodes~\cite{gros2021reconfigurable, Cui2022demo}, varactors~\cite{araghi2022reconfigurable}, electrically controlled microelectromechanical systems (MEMS), and liquid crystals~\cite{zhang2010voltage}, which makes \ac{RIS}-assisted systems energy-efficient. 
Among these hardware choices, PIN diodes have become the most prevailing tunable components, and have been widely applied to \acp{RIS} due to their ability to serve as high-speed microwave switches with low insertion loss and low control voltage~\cite{ismail2011investigation}. 
It is worth noting that, previous studies have pointed out the importance of considering the {\it dynamic power consumption} while constructing the power model for PIN diodes~\cite{wang2022reconfigurable}. 
Specifically, each PIN diode consumes a typical power of around $10\,{\rm mW}$ when it is ON, and the power consumption varies based on its configuration~\cite{wang2022reconfigurable, Huang2018energy, Mendez2016hybrid}. 
For an RIS equipped with 512 elements, by assuming half of the PIN diodes are ON, the power dissipation will be $2.56\,{\rm W}$. 
Compared to less than $10\,{\rm W}$ for \ac{BS} transmit power and $10\,{\rm mW}$ for each user~\cite{Huang2019recon}, the power dissipation of PIN diodes accounts for a significant proportion of the total power consumption in \ac{RIS}-assisted communication systems. 
Therefore, when optimizing \ac{RIS} configurations to meet energy-saving requirements in RIS-assisted systems, accurately modeling the power consumption attributed to PIN diodes is a crucial prerequisite.

\subsection{Prior Works} \label{Prior Works}
Traditional communication performance indicator is the {\it spectral efficiency} (SE). The SE optimization of RIS-assisted systems has been extensively studied, and various optimization schemes have been proposed~\cite{Boyd2004convex, Yuan2021intelli}. 
To further reduce power consumption of RIS-assisted systems, a comprehensive performance indicator called the {\it energy efficiency} (EE) has been studied in the literature~\cite{Wu2017an, Mahapatra2016energy}.
The EE is defined as the ratio of SE to total power consumption. 
Thus, different from SE optimization, EE optimization is generally more complicated due to the additional fractional structure. 
Therefore, SE optimization algorithms cannot be directly applied to EE optimization problems. 
To address the fractional structure that appears in the objective function of EE, effective algorithms such as sequential fractional programming (SFP) method~\cite{Huang2019recon} and quadratic transformation method~\cite{You2021energy} have been proposed. 

However, the power consumption model used in EE optimization in previous works is highly inaccurate. 
Most of the existing EE optimization algorithms assume the RIS power dissipation to be constant, i.e., independent of RIS configurations~\cite{Huang2019recon, You2021energy}. 
As mentioned earlier, for real-world phase-tuning components, the power caused by PIN diodes occupies a considerable portion of the total power consumption in RIS-assisted communication systems.  
Furthermore, the power consumption of PIN diodes in RIS varies significantly when they are configured to different states. 
Specifically, when configured to the ON state, i.e., the equivalent microwave switches admit the microwave signals, the diodes become extremely more power-hungry than in the OFF states~\cite{wang2022reconfigurable, Mendez2016hybrid}. 
Consequently, although the power consumption model employed in prior studies simplifies the algorithm design, it introduces severe inaccuracy for actual PIN diode-controlled RIS elements, leading to high additional power consumption when designing algorithms to optimize EE. 
This issue becomes more pronounced in scenarios involving a large number of RIS elements~\cite{Arun2020rfocus}, for instance, 1100~\cite{Pei2021ris} and 2304 elements~\cite{Cui2022demo}. 

Thus, if we fail to consider the impact of the ON-OFF power difference, it can result in a significant increase in power consumption and a substantial reduction in EE, especially when the number of RIS elements increases. 
In order to acquire a high EE, an effective design of the ON-OFF states of RIS elements is required to achieve high beamforming gain with fewer ON-state elements. 
Prior to designing algorithms for optimal EE, it is necessary to introduce an actual model of RIS power consumption with the consideration of the ON-OFF power difference. 
Unfortunately, existing works mentioned above have neglected this crucial point, leading to a severe deviation from realistic scenarios. 
Therefore, {\it how to accurately model the RIS power consumption and design its configuration} is a critical aspect of achieving optimal EE in RIS-assisted communication systems. 

\subsection{Our Works} \label{Our Works}

In this paper, we employ a more realistic power model for RIS-assisted systems, based on which we propose an effective algorithm to obtain optimal EE\footnote{Simulation codes will be provided to reproduce the results in this paper: \url{http://oa.ee.tsinghua.edu.cn/dailinglong/publications/publications.html}.}. Specifically, the contributions of this paper are summarized as follows: 
\begin{itemize}
    \item First, we introduce a realistic power dissipation model for downlink 1-bit RIS-assisted multi-user multiple-input single-output (MU-MISO) communication systems, which models the ON-OFF power difference of each RIS element. The proposed model better fits the actual RIS-assisted communication systems. Based on this model, we re-formulate the EE optimization problem. 
    \item Next, to solve the formulated EE optimization problem, we adopt an alternating optimization (AO) algorithm framework to optimize the \ac{BS} and RIS precoder separately. However, obtaining the RIS precoder is NP-hard due to the non-convex mixed-integer property. To solve this problem, we apply two different methods based on maximum gradient search and SDP relaxation, respectively. The former method has lower computational complexity, while the latter one achieves better performance. 
    \item Finally, we analyze the convergence and computational complexity of the proposed algorithms. Analysis results reveal that the exponential complexity of the original problem has been reduced to polynomial complexity. Simulation results verify the effectiveness of the proposed AO algorithmic framework with both methods, leading to a significant EE improvement in various scenarios. 
\end{itemize}

\subsection{Organization and Notation}
\label{Organization and Notation}

\textit{Organization:}
The paper is structured as follows. 
In Section~\ref{System Model}, we establish the signal model and the definition of EE for downlink RIS-assisted MU-MISO systems, and then formulate the more accurate EE optimization problem with the consideration of RIS element ON-OFF power difference. 
In Section~\ref{Algorithms}, we introduce the AO algorithm framework to decouple the original problem into two subproblems, i.e., the power allocation problem, and the RIS analog beamforming problem. 
The analytical solution to the power allocation problem and the analysis of the computational complexity and convergence are also discussed. 
In Section~\ref{RIS Analog Beamforming}, we focus on the non-convex mixed-integer RIS beamforming subproblem and provide two effective methods to acquire the near-optimal solution. The complexity and convergence analysis are also provided.  
In Section~\ref{Simulation Results}, simulation results are provided to verify the performance and effectiveness of the proposed algorithm.  
Section~\ref{Conclusion} concludes this paper. 

\textit{Notation:}
$\mathbb R$ and $\mathbb C$ represent the sets of real and complex numbers, respectively.
$\bm A^*$, $\bm A^{-1}$, $\bm A\T$, and $\bm A^{\rm H}$ indicate the conjugate, inverse, transpose, and conjugate transpose of matrix $\bm A$, respectively.
$\mathcal{CN}\left(\mu, \sigma^2 \right)$ refers to the complex univariate Gaussian distribution with mean $\mu$ and variance $\sigma^2$.
$\Vert \cdot\Vert_{n}$ denotes the $\mathcal{L}_{n}$-norm of its argument, respectively.
$\text{diag}(\cdot )$ is the diagonal operation. 
$\bm{1}_{M \times N}$ and $\bm{0}_{M \times N}$ are $M \times N$ matrices with all elements equal to $1$ and $0$, respectively. 
$\tr\left(\bm{X}\right)$ refers to the trace of the matrix $\bm{X}$.
$\bm{X} \succeq 0$ denotes a positive semi-definite matrix. 
$\otimes$, $\odot$ denote the Kronecker and Hadamard product of two matrices respectively. 
$\simeq$ indicates the equivalence of computational complexity order.

\section{System Model}
\label{System Model}
In this section, we will first specify the signal model in Subsection~\ref{Signal Model}. 
Then, a more realistic power model and EE will be introduced in Subsection~\ref{Energy Efficiency} with the consideration of ON-OFF power difference. 
Based on this, the EE optimization problem is formulated in Subsection~\ref{Problem Formulation}. 

\subsection{Signal Model}
\label{Signal Model}
We consider a downlink RIS-assisted MU-MISO system, where $K$ single-antenna users are served by an $M$-antenna \ac{BS}. 
The direct \ac{BS}-user link is assumed to be blocked as shown in Fig.~\ref{fig:System Model}. 
To ensure signal coverage and user experience, communication from BS to users is assisted by a RIS.
The RIS comprises $N_1$ reflecting elements in the horizontal direction and $N_2$ reflecting elements in the vertical direction, resulting in a total of $N = N_1 \times N_2$ reflecting elements.
Each RIS element is assumed to be binary-controlled, i.e., the diagonal phase-shift matrix ${\bm{\Theta}}$ of the RIS only takes two possible values 
\begin{equation}
    \bm{\Theta} = \diag(e^{\j \bm{\theta}}) = \diag\left( \left[ e^{\j \theta_1}, e^{\j \theta_2}, ..., e^{\j \theta_N} \right] \right), \ \theta_n \in \{ 0, \pi \}. 
\end{equation}

\ifx\onecol\undefined
\begin{figure}[!t]
    \centering
    \myincludegraphics{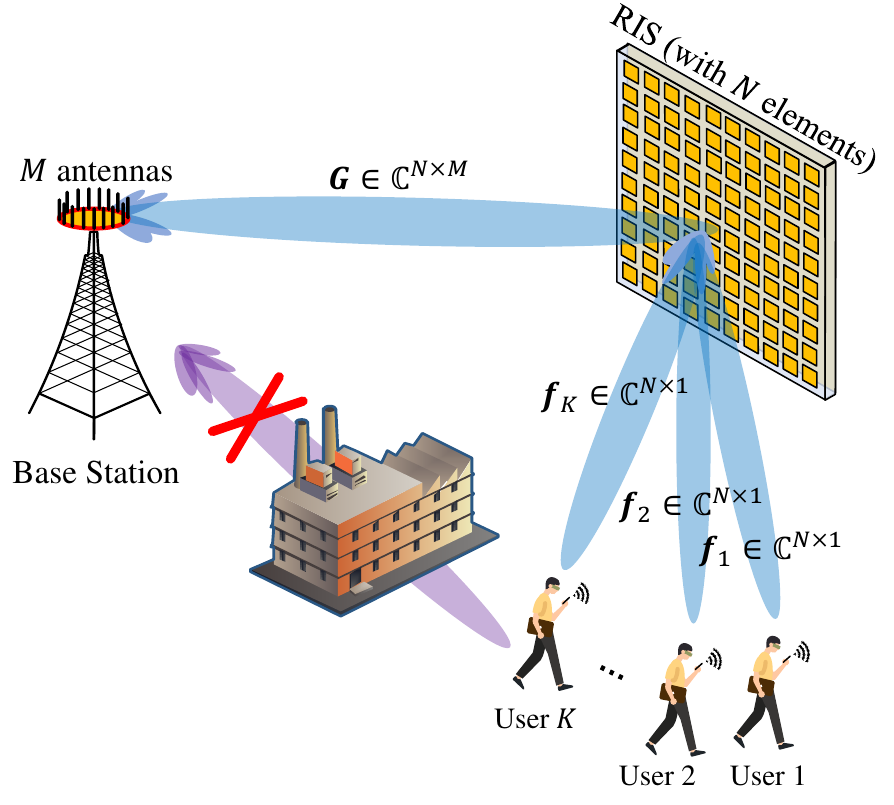}
    \caption{RIS-assisted MU-MISO downlink system.}
    \label{fig:System Model}
\end{figure}
\else
\begin{figure}[!t]
    \centering
    \myincludegraphics{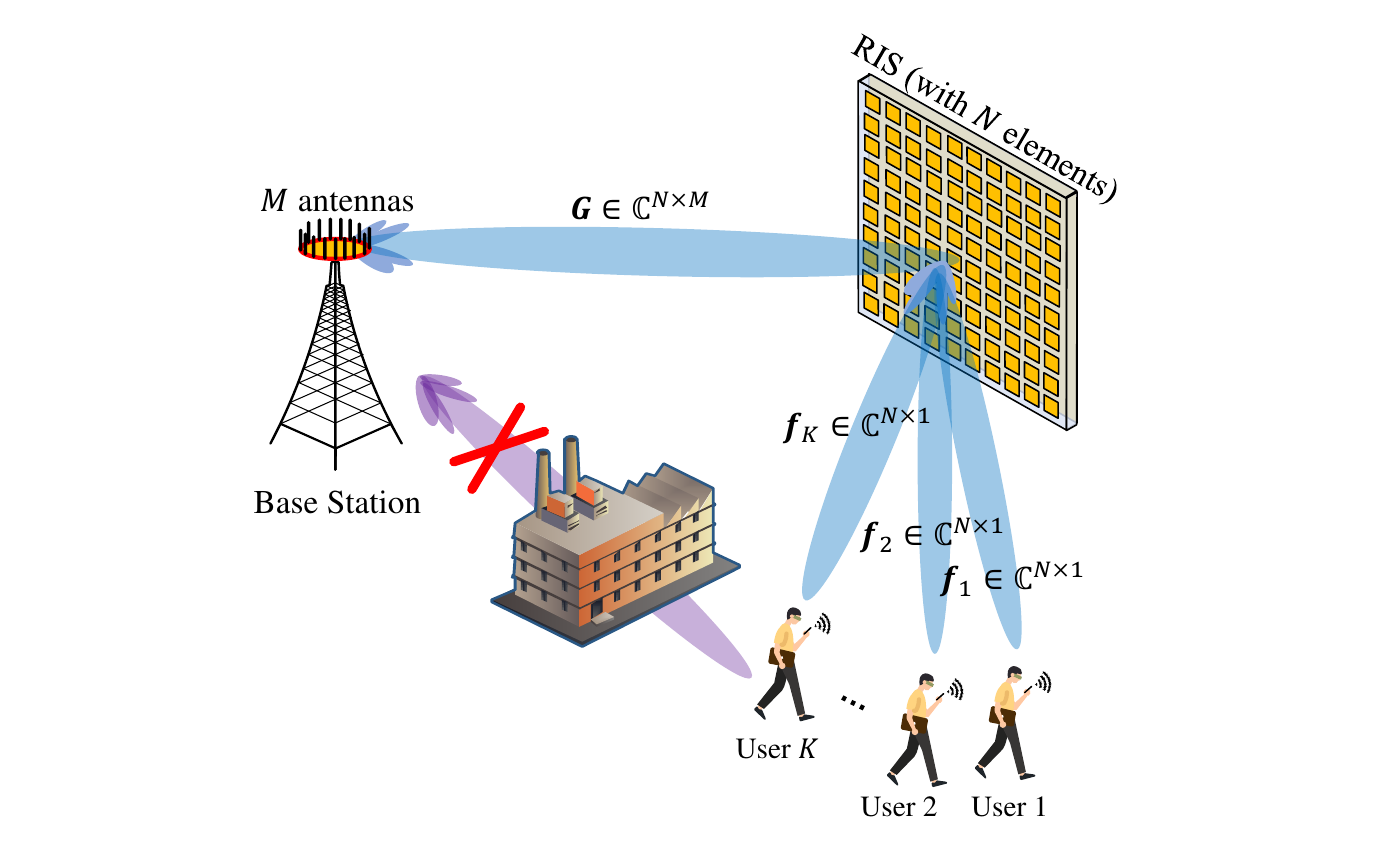}
    \caption{RIS-assisted MU-MISO downlink system.}
    \label{fig:System Model}
\end{figure}
\fi

The RIS-BS channel is denoted as $\bm{G} \in \mathbb{C}^{N \times M}$, and the channel from RIS to the $k$-th user is denoted as $\bm{f}_k\H \in \mathbb{C}^{1 \times N}$. 
Then, the signal received by the users can be represented as 
\begin{equation}
    \bm{y} = \bm{F}\H \bm{\Theta G W s} + \bm{n}, 
\end{equation}
where $\bm{F} = [\bm{f}_1, \bm{f}_2, ..., \bm{f}_K]$ represents the equivalent channel from RIS to each user, $\bm{s} = [s_1, s_2, ..., s_K]\T$ represents the signal transmitted to each user, $\bm{W}$ represents the digital precoding from BS with the power constraint $\tr(\bm{W}\H \bm{W}) \leq \Pmax$, and $\bm{n} \sim \mathcal{CN}({\bm 0}, \sigma_n^2{\bm I}_K) $ represents the \ac{AWGN} imposed at each receiver.

\subsection{Energy Efficiency}
\label{Energy Efficiency}
The total power consumption can be modeled as follows 
\begin{equation}
    P_{\text{all}} = P_{\text{static}} + P_{\text{RIS}} + \nu^{-1} P_{\text{transmit}}, 
\end{equation}
in which $P_{\text{static}}$ represents the overall static power consumption at BS, users, and RIS, $P_{\text{transmit}} = \tr\left( \bm{W}\H\bm{W} \right)$ represents the BS transmit power, and 
$\nu$ represents the efficiency of the power transmit amplifiers, which is considered as $1$ in the following text. 
It is noteworthy that the power consumption of RIS elements is considered as a fixed value in most of the prior works~\cite{Huang2019recon, You2021energy, Huang2018energy}, which is not in accord with the actual situation discussed in Section~\ref{Introduction}.  
Thus, a binary-control RIS is considered here, with the difference in ON-OFF power consumption of the PIN diode on each RIS element~\cite{wang2022reconfigurable}, i.e.,
\begin{equation}
    P_{\text{RIS}} = \left\Vert \bm{\theta} \right\Vert_0 P_0, \ \theta_n \in \{ 0, \pi \}, 
\end{equation}
where $P_0$ is the power dissipation of each RIS element when the corresponding PIN diode is turned ON\footnote{In this paper, we assume that the reflection coefficient of each RIS element is tuned by only one PIN diode. } by applying a bias current~\cite{gros2021reconfigurable}, and the $n$-th element $\theta_n$ of $\bm{\theta}$ is the phase-shift configuration of the $n$-th RIS element. 
With the discussions above, the SE and EE can be written as follows~\cite{Wen2011onthe}, 
\ifx\onecol\undefined
\begin{align}
    \SE \left( \bm{\Theta}, \bm{W} \right) &=  \sum_{k=1}^K \log_2 \left( 1 + \frac{\left\vert \bm{f}_k\H \bm{\Theta G} \bm{w}_k \right\vert^2}{\sum_{k' \neq k} \left\vert \bm{f}_k\H \bm{\Theta G} \bm{w}_{k'} \right\vert^2 + \sigma_n^2 } \right), \label{equ: SE definition} \\ 
    \EE \left( \bm{\Theta}, \bm{W} \right) &= \frac{\text{BW} \times \SE \left( \bm{\Theta, \bm{W}} \right)}{P_{\text{static}} + P_0 \left\Vert \bm{\theta} \right\Vert_0 + \tr\left(\bm{W}\H\bm{W}\right) }. \label{equ: EE definition}
\end{align}
\else
\begin{subequations}
    \begin{align}    
    & \SE \left( \bm{\Theta}, \bm{W} \right) = \sum_{k=1}^K \log_2 \left( 1 + \frac{\left\vert \bm{f}_k\H \bm{\Theta G} \bm{w}_k \right\vert^2}{\sum_{k' \neq k} \left\vert \bm{f}_k\H \bm{\Theta G} \bm{w}_{k'} \right\vert^2 + \sigma_n^2 } \right) \ [\text{bits/s/Hz}], \\
    & \EE \left( \bm{\Theta}, \bm{W} \right) = \frac{\text{BW} \times \SE \left( \bm{\Theta, \bm{W}} \right)}{P_{\text{static}} + P_0 \left\Vert \bm{\theta} \right\Vert_0 + \tr\left(\bm{W}\H\bm{W}\right) } \ [\text{bits/Joule}]. 
    \end{align}
\end{subequations}
\fi

\subsection{Problem Formulation}
\label{Problem Formulation}
In this paper, our target is to acquire the maximum EE by jointly designing the digital beamforming matrix $\bm{W}$ and the RIS analog beamforming vector $\bm{\theta}$, which can be expressed as
\begin{subequations}
    \begin{align}
    \mathcal{P}_0: \max_{\bm{\Theta}, \bm{W}} \ &\EE \left( \bm{\Theta}, \bm{W} \right), \label{pro:0-a} \\
    \text{s.t.} \ &\tr\left(\bm{W}\H \bm{W}\right) \leq \Pmax, \label{pro:0-b} \\
    & \SE_{k} \geq \SE_{\text{min}}, \ \forall k \in \mathcal{K}, \label{pro:0-c} \\
    & \theta_n \in \{0, \pi\}, \ \forall n \in \mathcal{N} \label{pro:0-d} . 
    \end{align}
\end{subequations}
As shown in \eqref{pro:0-c}, we set a minimum spectrum efficiency value $\SE_{\text{min}}$ for each user as a basic communication requirement. 

The EE optimization problem $\mathcal{P}_0$ is a non-concave mixed integer programming due to the non-convex target function \eqref{pro:0-a} and the sparse constraint \eqref{pro:0-d}, which makes the problem extremely difficult and complicated. 
In the next part of this paper, we will develop an effective optimization algorithm with an acceptable computational complexity.

\section{Algorithms}
\label{Algorithms}
In this section, we will present an algorithm to address the EE optimization problem $\mathcal{P}_0$. 
Firstly, we will introduce an alternating optimization procedure in order to acquire the optimal phase-shifts $\bm{\theta}$ and BS precoders $\bm{W}$ in Subsection~\ref{Alternating Optimization}. 
Then, the solution of the power allocation problem will be discussed in Subsection~\ref{Solution of the Power Allocation Problem}, i.e. the optimal $\bm{W}$ with fixed $\bm{\theta}$. 
The convergence and the computational complexity of the algorithm will be discussed in Subsection~\ref{Complexity & Convergence Analysis 1}. 
For clarity, the algorithm to solve the RIS analog beamforming problem will be designed in the next section, whose difficulty and complexity lead to a separate section to discuss. 

\subsection{Alternating Optimization}
\label{Alternating Optimization}
In order to fully eliminate the interference of signal in different channels, the Zero-Forcing (ZF) digital precoder~\cite{Peel2005Avector} can be utilized here to obtain a feasible solution, i.e.,  
\begin{equation}
    \bm{W} = \bm{H} \left( \bm{H}\H \bm{H} \right)^{-1} \bm{P}^{\frac{1}{2}}, 
\end{equation}
where $\bm{H}\H = \bm{F}\H \bm{\Theta G}$ represents the cascade channel from BS to users, the diagonal matrix $\bm{P}$ represents the power allocation of each user, whose $k$-th diagonal element $p_k$ represents the signal power received by the $k$-th user. 
With the ZF precoder, the signal received by users can be rewritten as follows: 
\begin{equation}
    \bm{y} = \bm{P}^{\frac{1}{2}} \bm{s} + \bm{n},  
\end{equation}
and the corresponding transmit power constraints~\eqref{pro:0-b} of $\mathcal{P}_0$ can be rewritten as 
\begin{equation}
    \tr \left( \bm{W}\H \bm{W} \right) = \tr \left( \bm{P}^{\frac{1}{2}} \left( \bm{H}\H \bm{H} \right)^{-1} \bm{P}^{\frac{1}{2}} \right) \leq P_{\text{max}}.
\end{equation}
Thus, the spectrum efficiency can also be rewritten as 
\begin{equation}
    \SE\left( \bm{\Theta}, \bm{P} \right) = \SE\left( \bm{P} \right) = \sum_{k=1}^K \log_2 \left( 1 + \frac{p_k}{\sigma^2} \right),
\end{equation}
where $p_k$ represents the $k$-th diagonal element of $\bm{P}$, i.e. the received signal power of the $k$-th user. 
Then, the problem $\mathcal{P}_0$ can be expressed as 
\ifx\onecol\undefined
\begin{subequations}
    \begin{align}
    \mathcal{P}_0': \max_{\bm{\Theta}, \bm{P}} \ & \frac{\text{BW} \sum_{k=1}^K \log_2 \left( 1 + \frac{p_k}{\sigma_n^2} \right)}{P_{\text{static}} + P_0 \left\Vert \bm{\theta} \right\Vert_0 + \tr \left( \bm{P}^{\frac{1}{2}} \left( \bm{H}\H \bm{H} \right)^{-1} \bm{P}^{\frac{1}{2}} \right)},  \\
    \text{s.t.} \ & \tr \left( \bm{P}^{\frac{1}{2}} \left(\bm{H}\H \bm{H}\right)^{-1} \bm{P}^{\frac{1}{2}} \right) \leq \Pmax, \\
    & p_k \geq p_{\text{min}}, \ \forall k \in \mathcal{K} \\
    & \theta_n \in \{ 0, \pi \}, \ \forall n \in \mathcal{N}, 
    \end{align}
\end{subequations}
\else
\begin{subequations}
    \begin{align}
    \mathcal{P}_0': \max_{\bm{\Theta}, \bm{P}} \ & \frac{\text{BW}}{P_{\text{static}} + P_0 \left\Vert \bm{\theta} \right\Vert_0 + \tr \left( \bm{P}^{\frac{1}{2}} \left( \bm{H}\H \bm{H} \right)^{-1} \bm{P}^{\frac{1}{2}} \right)} \sum_{k=1}^K \log_2 \left( 1 + \frac{p_k}{\sigma_n^2} \right), \\
    \text{s.t.} \ & \tr \left( \bm{P}^{\frac{1}{2}} \left(\bm{H}\H \bm{H}\right)^{-1} \bm{P}^{\frac{1}{2}} \right) \leq \Pmax, \\
    & p_k \geq p_{\text{min}}, \ \forall k \in \mathcal{K} \\
    & \theta_n \in \{ 0, \pi \}, \ \forall n \in \mathcal{N}, 
    \end{align}
\end{subequations}
\fi
where $p_{\text{min}} = \sigma^2 \left( 2^{\SE_{\text{min}}} - 1 \right)$ represents the minimum received power requirement of each user in order to ensure the basic communication spectrum efficiency $\SE_{\text{min}}$. 

With the discussions above, the original optimization problem ${\mathcal P}_0$ is transformed to ${\mathcal P}_0'$ with the optimization variables $\bm{\Theta}$ and $\bm{P}$. 
Considering the difficulty to acquire optimal $\bm{\Theta}$ and $\bm{P}$ simultaneously, an alternating optimization (AO) algorithm can be applied to solve the problem. 
Firstly, the power allocation matrix $\bm{P}$ can be optimized with fixed $\bm{\Theta}$, and then the RIS analog beamforming matrix $\bm{\Theta}$ can be optimized with fixed $\bm{P}$. 
Thus, $\bm{P}$ and $\bm{\Theta}$ can be updated until convergence. 

\subsection{Solution of the Power Allocation Problem}
\label{Solution of the Power Allocation Problem}
According to the AO algorithm of problem $\mathcal{P}_0'$, the first step is to find the optimal power allocation matrix $\bm{P}$ with the fixed RIS configuration $\bm{\Theta}$, which leads to a power allocation problem as follows: 
\begin{subequations}
    \begin{align}
    \mathcal{P}_1: \max_{\bm{P}} \ & \frac{1}{P_1 + \sum_{k=1}^K p_k t_k} \sum_{k=1}^K \log \left( 1 + \frac{p_k}{\sigma_n^2} \right), \label{pro:1-a} \\
    \text{s.t.} \ & \sum_{k=1}^K p_k t_k \leq \Pmax, \label{pro:1-b} \\
    & p_k \geq p_{\text{min}}, \label{pro:1-c} 
    \end{align}
\end{subequations}
where $P_1 \triangleq P_{\text{static}} + P_0 \left\Vert \bm{\theta} \right\Vert_0$ and $t_k$ is the $k$-th diagonal element of $\left( \bm{H}\H \bm{H} \right)^{-1}$. 
Although the constraints \eqref{pro:1-b} and \eqref{pro:1-c} are both affine, the problem $\mathcal{P}_1$ is also a thorny matter due to the non-concave target function \eqref{pro:1-a}. 
One feasible method is to introduce a relaxation variable in order to convert fractions to polynomials. 
Thus, the problem above can be solved by Dinkelbach's method as follows\cite{Huang2019recon}: 
\ifx\onecol\undefined 
\begin{subequations}
    \begin{equation}
        \begin{aligned}
            \bm{P}^{(i)} = \arg \max_{\bm{P}} & \ \sum_{k=1}^K \log_2 \left( 1 + \frac{p_k}{\sigma_n^2} \right) \\
            & - \lambda^{(i-1)} \left( P_1 + \sum_{k=1}^K p_k t_k \right), \\
            \text{s.t.} & \sum_{k=1}^K p_k t_k \leq \Pmax, \ p_k \geq p_{\text{min}}
            \label{equ:Dinkelbach1} 
        \end{aligned}
    \end{equation}
    \begin{equation}
        \lambda^{(i)} = \frac{1}{P_1 + \sum_{k=1}^K p_k^{(i)} t_k} \sum_{k=1}^K \log_2 \left( 1 + \frac{p_k^{(i)}}{\sigma_n^2} \right). \label{equ:Dinkelbach2}
    \end{equation}
\end{subequations}
\else
\begin{subequations}
    \begin{align}
    & \bm{P}^{(i)} = \arg \max_{\bm{P}} \ \sum_{k=1}^K \log_2 \left( 1 + \frac{p_k}{\sigma_n^2} \right) - \lambda^{(i-1)} \left( P_1 + \sum_{k=1}^K p_k t_k \right), \ \text{s.t.} \ \sum_{k=1}^K p_k t_k \leq \Pmax, \ p_k \geq p_{\text{min}} \label{equ:Dinkelbach1} \\ 
    & \lambda^{(i)} = \frac{1}{P_1 + \sum_{k=1}^K p_k^{(i)} t_k} \sum_{k=1}^K \log_2 \left( 1 + \frac{p_k^{(i)}}{\sigma_n^2} \right). \label{equ:Dinkelbach2}
    \end{align}
\end{subequations}
\fi
The optimization problem~\eqref{equ:Dinkelbach1} is convex, whose analytical solution will be given in \textbf{Appendix~\ref{Analytical solution of the problem Dinkelbach1}}. 
Then, the procedures to solve $\mathcal{P}_1$ can be summarized in \textbf{Algorithm~\ref{alg:power allocation problem}}. 
\ifx\onecol\undefined
\begin{algorithm}
    \caption{Power Allocation Problem}
    \label{alg:power allocation problem}
    \setstretch{1.35}
    \begin{algorithmic}[1]
        \REQUIRE Numbers of users $K$; Power fading coefficients $t_1, ..., t_K$; Variance of the \ac{AWGN} $\sigma_n^2$. 
        \ENSURE Power allocatin matrix $\bm{P}={\rm diag}(p_1, p_2, \cdots, p_K)$. 
        \STATE Find $\zeta$ such that: $\sum_k \left\{ \zeta - t_k \sigma_n^2, t_k p_{\text{min}} \right\} = \Pmax$
        \FOR {$i = 1, 2, ..., N_{\text{iter}}$}
            \STATE $\xi^{(i)} \leftarrow \min\left\{ \zeta, 1/\left( \lambda^{(i-1)} \log 2 \right) \right\}$. 
            \STATE $p_k^{(i)} \leftarrow \max\left\{ \left( \xi - t_k \sigma_n^2 \right)/t_k, \Pmin \right\}$.
            \STATE $\lambda^{(i)} \leftarrow \sum_k \log_2 \left\{ 1+p_k^{(i)}/\sigma_n^2 \right\} / \left( P_1 + \sum_k p_kt_k \right)$. 
        \ENDFOR
        \RETURN Optimized $\bm{P}$ 
    \end{algorithmic}
\end{algorithm}
\else
\begin{algorithm}
    \caption{Power Allocation Problem}
    \label{alg:power allocation problem}
    \setstretch{1.35}
    \begin{algorithmic}[1]
        \REQUIRE Numbers of users $K$; Power fading coefficients $t_1, ..., t_K$; Variance of the \ac{AWGN} $\sigma_n^2$. 
        \ENSURE Power allocatin matrix $\bm{P}={\rm diag}(p_1, p_2, \cdots, p_K)$. 
        \STATE Find $\zeta$ such that: $\sum_{k=1}^K \left\{ \zeta - t_k \sigma_n^2, t_k p_{\text{min}} \right\} = \Pmax$
        \FOR {$i = 1, 2, ..., N_{\text{iter}}$}
            \STATE $\xi^{(i)} \leftarrow \min\left\{ \zeta, 1/\left( \lambda^{(i-1)} \log 2 \right) \right\}$. 
            \STATE $p_k^{(i)} \leftarrow \max\left\{ \left( \xi - t_k \sigma_n^2 \right)/t_k, \Pmin \right\}$.
            \STATE $\lambda^{(i)} \leftarrow \sum_{k=1}^K \log_2 \left\{ 1+p_k^{(i)}/\sigma_n^2 \right\} / \left( P_1 + \sum_{k=1}^K p_kt_k \right)$. 
        \ENDFOR
        \RETURN Optimized $\bm{P}$ 
    \end{algorithmic}
\end{algorithm}
\fi

\vspace{-2mm}
\subsection{Complexity \& Convergence Analysis}
\label{Complexity & Convergence Analysis 1}
Firstly, the computational complexity will be derived as follows. 
According to \textbf{Algorithm \ref{alg:power allocation problem}}, the complexity of the power allocation problem mainly comes from the iteration step. 
For each step, the values of $\xi$, $p_k$, and $\lambda$ are calculated in turn, which leads to a linear complexity $\mathcal{O}(K)$. 
Thus, the computational complexity of the power allocation problem is $\mathcal{O}(N_{\rm iter}K)$, where $N_{\rm iter}$ represents the number of iterations of the algorithm. 
The computational complexity of the RIS beamforming problem will be discussed in Subsection \ref{Complexity & Convergence Analysis 2}. 

We focus on the convergence of the power allocation problem. 
According to the optimality of $p_k^{(i)}$ in~\eqref{equ:Dinkelbach1}, we have 
\ifx\onecol\undefined
\begin{equation}
    \begin{aligned}
    \mathcal{P} \left( p_k^{(i)} \right) = & \sum_{k=1}^K \log_2 \left( 1 + \frac{p_k^{(i)}}{\sigma_n^2} \right) - \lambda^{(i-1)} \left( P_1 + \sum_{k=1}^K p_k^{(i)} t_k \right) \\ 
    \geq & \ \mathcal{P} \left( p_k^{(i-1)} \right) = \sum_{k=1}^K \log_2 \left( 1 + \frac{p_k^{(i-1)}}{\sigma_n^2} \right) \\ 
    & - \lambda^{(i-1)} \left( P_1 + \sum_{k=1}^K p_k^{(i-1)} t_k \right) = 0. 
    \end{aligned}
\end{equation}
\else
\begin{equation}
    \begin{aligned}
    & \mathcal{P} \left( p_k^{(i)} \right) = \sum_{k=1}^K \log_2 \left( 1 + \frac{p_k^{(i)}}{\sigma_n^2} \right) - \lambda^{(i-1)} \left( P_1 + \sum_{k=1}^K p_k^{(i)} t_k \right) \\ 
    \geq \ & \mathcal{P} \left( p_k^{(i-1)} \right) = \sum_{k=1}^K \log_2 \left( 1 + \frac{p_k^{(i-1)}}{\sigma_n^2} \right) - \lambda^{(i-1)} \left( P_1 + \sum_{k=1}^K p_k^{(i-1)} t_k \right) = 0. 
    \end{aligned}
\end{equation}
\fi
Therefore, it is obvious that $\lambda^{(i)} \geq \lambda^{(i-1)}$, which proves the convergence of $\mathcal{P}_1$. 

As for the convergence of AO algorithm, consider the $j$-th iteration where $\EE^{(j)}_1$ denotes the EE after power allocation (optimizing $\bm{P}$ with fixed $\bm{\Theta}$) and $\EE^{(j)}_2$ denotes the EE after RIS beamforming (optimizing $\bm{\Theta}$ with fixed $\bm{P}$). 
With the discussion about the convergence of $\mathcal{P}_1$, it is ensured that $\EE^{(j)}_1 \geq \EE^{(j-1)}_2$. 
As long as the RIS beamforming problem can be solved effectively, i.e. the algorithm of the problem satisfies $\EE^{(j)}_2 \geq \EE^{(j)}_1$, which will be explained in detail in Section \ref{RIS Analog Beamforming}, we will have $\EE^{(j)}_2 \geq \EE^{(j)}_1 \geq \EE^{(j-1)}_2$. 
This proves the convergence of the AO algorithm.

\section{RIS Analog Beamforming}
\label{RIS Analog Beamforming}
In this section, we focus on the discussion about the algorithm of RIS analog beamforming. 
According to the AO algorithm in Section \ref{Algorithms}, the next step is to optimize $\bm{\Theta}$ with fixed $\bm{P}$, which can be expressed as follows, 
\begin{subequations}
    \begin{align}
    \mathcal{P}_2: \min_{\bm{\Theta}} \ & P_0 \left\Vert \bm{\theta} \right\Vert_0 + \sum_{k=1}^K p_k t_k, \label{pro:2-a} \\
    \text{s.t.} \ & \sum_{k=1}^K p_k t_k \leq \Pmax, \label{pro:2-b} \\
    & \theta_n \in \{ 0, \pi \}, \ \forall n \in \mathcal{N}. \label{pro:2-c}
    \end{align}
\end{subequations}
We denote $\bm{q} = e^{\j \bm{\theta}}$. Since $\theta_n \in \{0, \pi\}$, we have $q_n\in \{-1, 1\}$ and $\left\Vert \bm{\theta} \right\Vert_0 = -\frac{1}{2}\bm{1}\T \bm{q}+N/2$. Then $\mathcal{P}_2$ can be rewritten as 
\begin{subequations}
    \begin{align}
    \mathcal{P}_2': \min_{\bm{q}} \ & -\frac{1}{2} P_0 \bm{1}\T \bm{q} + \sum_{k=1}^K p_k t_k, \label{pro:2'-a} \\ 
    \text{s.t.} \ & \sum_{k=1}^K p_k t_k \leq \Pmax, \label{pro:2'-b} \\
    & q_n \in \{ -1, 1 \}, \ \forall n \in \mathcal{N}. \label{pro:2'-c}
    \end{align}
\end{subequations}
It should be noted that $t_k$ is a function of $\bm{q}$ in $\mathcal{P}_2'$. 
As we have mentioned, the non-convexity of the target function~\eqref{pro:2-a} and the discrete constraint~\eqref{pro:2-c} makes $\mathcal{P}_2$ difficult to solve. 
Generally, it is an NP-hard problem due to the integer constraint~\eqref{pro:2-c}, so it is almost impossible to acquire an optimal solution with low computational time. 
However, some effective methods such as heuristic search and proper relaxations are helpful to obtain a computationally feasible solution. 
In Subsection~\ref{Search with the Maximum Gradient} we propose an algorithm based on maximum gradient search, which has a low computational cost and thus can solve the problem efficiently. 
Due to the fact that gradient-based algorithms may converge to local optimum in some situations, an alternative algorithm based on SDP relaxation is proposed in Subsection~\ref{SDP Relaxation}, and the appropriate solution of problem $\mathcal{P}_2$ will be given. 
The computational complexity and convergence will be analyzed in Subsection~\ref{Complexity & Convergence Analysis 2}. 

\subsection{Search with the Maximum Gradient}
\label{Search with the Maximum Gradient}
One of the most popular solutions to mixed integer programming $\mathcal{P}_2'$ is based on searching methods, such as the Branch and Bound method. 
The fatal defect of this type of method is that it usually has unacceptable computational complexity, e.g., $\mathcal{O}(2^N)$. 
Although gradient descent methods in continuous variable optimization cannot be applied to discrete cases directly, they still provide some insights. 
In discrete cases, the direction with the maximum gradient value can also be regarded as the fastest direction to make the objective function decline. 
Then, we can design the searching method with the maximum gradient. 

Firstly, the gradient of the objective function in \eqref{pro:2'-a} (defined as $g(\bm{q})$) can be expressed as 
\ifx\onecol\undefined
\begin{equation}
    \begin{aligned}
    \frac{\partial g(\bm{q})}{\partial q_n} 
    & = - \frac{1}{2} P_0 + \sum_{k=1}^K p_k \left[ \frac{\partial \left( \bm{H}\H \bm{H} \right)^{-1}}{\partial q_n} \right]_{(k, k)} \\ 
    & = - \frac{1}{2} P_0 - \sum_{k=1}^K p_k \left[ \left( \bm{H}\H \bm{H} \right)^{-1} \frac{\partial \bm{H}\H \bm{H}}{\partial q_n} \left( \bm{H}\H \bm{H} \right)^{-1} \right]_{(k, k)} \\ 
    & = - \frac{1}{2} P_0 - \sum_{k=1}^K p_k \left[ \left( \bm{H}\H \bm{H} \right)^{-1} \left( \bm{F}\H \diag(\bm{e}_n) \bm{G H} \right. \right. \\
    & \ \ \ \left. \left. + \bm{H}\H \bm{G}\H \diag(\bm{e}_n) \bm{F} \right) \left( \bm{H}\H \bm{H} \right)^{-1} \right]_{(k, k)}. 
    \end{aligned}
    \label{equ:gradient of g}
\end{equation}
\else
\begin{equation}
    \begin{aligned}
    \frac{\partial g(\bm{q})}{\partial q_n} & = - \frac{1}{2} P_0 + \sum_{k=1}^K p_k \left[ \frac{\partial \left( \bm{H}\H \bm{H} \right)^{-1}}{\partial q_n} \right]_{(k, k)} \\ 
    & = - \frac{1}{2} P_0 - \sum_{k=1}^K p_k \left[ \left( \bm{H}\H \bm{H} \right)^{-1} \frac{\partial \bm{H}\H \bm{H}}{\partial q_n} \left( \bm{H}\H \bm{H} \right)^{-1} \right]_{(k, k)} \\ 
    & = - \frac{1}{2} P_0 - \sum_{k=1}^K p_k \left[ \left( \bm{H}\H \bm{H} \right)^{-1} \left( \bm{F}\H \diag(\bm{e}_n) \bm{G H} + \bm{H}\H \bm{G}\H \diag(\bm{e}_n) \bm{F} \right) \left( \bm{H}\H \bm{H} \right)^{-1} \right]_{(k, k)}. 
    \end{aligned}
    \label{equ:gradient of g}
\end{equation}
\fi
Based on \eqref{equ:gradient of g}, we can design the searching method. 
Then, each $q_n$ changes from large $\partial g / \partial q_n \times q_n$ to small ones, during which the value of $g(\bm{q})$ and the feasibility of $\bm{q}$ will be verified. 
If the updated $\bm{q}^*$ is a feasible and better solution than $\bm{q}$, i.e. $g(\bm{q}^*) \leq g(\bm{q})$, the update will be kept and go on to the next. 
Details of the method are summarized in \textbf{Algorithm \ref{alg:search with gradient}}. 
\begin{algorithm}[H]
    \caption{Search with the Maximum Gradient}
    \label{alg:search with gradient}
    \setstretch{1.35}
    \begin{algorithmic}[1]
        \REQUIRE Number of RIS elements $N$; Channel matrix $\bm{F}, \bm{G}$; Energy consumption of each RIS element $P_{\text{RIS}}$; Initial RIS state $\bm{q}^{(0)}$; Ratio of each epoch $\rho$; Threshold $\varepsilon$. 
        \ENSURE RIS beamforming state $\bm{q}$.
        \WHILE {$ \left\Vert \bm{q}^{(i)} - \bm{q}^{(i-1)} \right\Vert_0 \geq \varepsilon $}
            \STATE Calculate the gradient of $g(\bm{q}^{(i-1)})$ according to \eqref{equ:gradient of g}. 
            \STATE Sort the product of $\bm{q}_n$ and $\partial g(\bm{q}) / \partial q_n$ in descending order $ \bm{d} $. 
            \STATE $\bm{q}^{(i)} \leftarrow \bm{q}^{(i-1)}$
            \FOR {$j = 1, 2, ..., \text{round}(\rho N)$}
                \STATE $\bm{q}^{(i)}_{d_j} \leftarrow - \bm{q}^{(i)}_{d_j}$
                \IF {This solution $\bm{q}^{(i)}$ is unfeasible or more energy consumption}
                    \STATE $\bm{q}^{(i)}_{d_j} \leftarrow - \bm{q}^{(i)}_{d_j}$
                \ENDIF
            \ENDFOR
            \STATE $i \leftarrow i + 1$
        \ENDWHILE
        \RETURN $\bm{q}$ 
    \end{algorithmic}
\end{algorithm}

\subsection{SDP Relaxation}
\label{SDP Relaxation}
In the previous subsection, the RIS phase-shift vector $\bm q$ is obtained by applying an effective heuristic search to the problem $\mathcal{P}_2'$ in {\bf Algorithm~\ref{alg:search with gradient}}, which does not guarantee the optimality. 
A more reasonable approach is to construct a solution to $\mathcal{P}_2'$ by analyzing the inherent mathematical structure of this optimization problem. 
Note that the constraints~\eqref{pro:2'-b} (or~\eqref{pro:2-b}) are applied to the trace of the inverse matrix of $\bm{H}\H \bm{H}$, which is further a quadratic function of ${\bm q}$ according to the definition of ${\bm H}$. Thus, if we replace ${\bm q}$ by a quadratic variable $\bm{X} \in \mathbb{R}^{(N+1)\times(N+1)}$ defined as 
\begin{equation}
    \bm{X} = 
    \begin{bmatrix}
        \bm{q}\bm{q}\T & \bm{q} \\ 
        \bm{q}\T & 1 
    \end{bmatrix}, 
\end{equation}
then the constraint~\eqref{pro:2'-b} can be expressed as $\tr({\bm Y}^{-1})$, where the elements of $\bm Y$ are linear combinations of the elements in $\bm X$. 
Since linear transform does not alter convexity, $\tr({\bm Y}^{-1})$ is a convex function of the new matrix argument $\bm X$. Through this change of variable, i.e., ${\bm q}\to{\bm X}$, the target function~\eqref{pro:2'-a} of $\mathcal{P}_2'$ is made convex simultaneously. Thus, $\mathcal{P}_2'$ can be rewritten as 
\ifx\onecol\undefined
\begin{subequations}
    \begin{align}
    \mathcal{P}_3: \min_{\bm{X}} & -\frac{1}{4}P_{\text{RIS}} \tr \left( \bm{E}_0 \bm{X} \right) + \tr \left(\bm{F}_0\H \left( \bm{X} \odot \bm{G}_0 \right) \bm{F}_0 \bm{P}^{-1}\right)^{-1}, \label{pro:3-a} \\ 
    \text{s.t.} \ & \tr\left( \left(\bm{F}_0\H \left( \bm{X} \odot \bm{G}_0 \right) \bm{F}_0 \bm{P}^{-1}\right)^{-1} \right) \leq \Pmax, \label{pro:3-b} \\ 
    & \tr \left( \bm{E}_{i, i} \bm{X} \right) = 1, \ i \in {1, 2, ..., N+1}, \label{pro:3-c} \\ 
    & \bm{X} \succeq 0, \label{pro:3-d} \\ 
    & \text{rank} \left( \bm{X} \right) = 1, \label{pro:3-e}
    \end{align}
\end{subequations}
\else
\begin{subequations}
    \begin{align}
    \mathcal{P}_3: \min_{\bm{X}} \ & -\frac{1}{4}P_{\text{RIS}} \tr \left( \bm{E}_0 \bm{X} \right) + \tr\left( \left(\bm{F}_0\H \left( \bm{X} \odot \bm{G}_0 \right) \bm{F}_0 \bm{P}^{-1}\right)^{-1} \right), \label{pro:3-a} \\ 
    \text{s.t.} \ & \tr\left( \left(\bm{F}_0\H \left( \bm{X} \odot \bm{G}_0 \right) \bm{F}_0 \bm{P}^{-1}\right)^{-1} \right) \leq \Pmax, \label{pro:3-b} \\ 
    & \tr \left( \bm{E}_{i, i} \bm{X} \right) = 1, \ i \in {1, 2, ..., N+1}, \label{pro:3-c} \\ 
    & \bm{X} \succeq 0, \label{pro:3-d} \\ 
    & \text{rank} \left( \bm{X} \right) = 1, \label{pro:3-e}
    \end{align}
\end{subequations}
\fi
where $\bm{E}_0 \in \mathbb{R}^{(N+1) \times (N+1)}$, $\bm{E}_{i, i} \in \mathbb{R}^{(N+1) \times (N+1)}$, $\bm{F}_0 \in \mathbb{C}^{(N+1) \times K}$, and $\bm{G}_0 \in \mathbb{R}^{(N+1) \times (N+1)}$ is defined as 
\ifx\onecol\undefined
\begin{equation}
    \begin{aligned}
        & \bm{E}_0 = 
        \begin{bmatrix}
            \bm{0}_{N \times N} & \bm{1}_{N \times 1} \\ \bm{1}_{1 \times N} & 0
        \end{bmatrix}, 
        \bm{E}_{i, i} = \diag(\bm{e}_i), \\ 
        & \bm{G}_0 = 
        \begin{bmatrix}
            \bm{G}\bm{G}\H & \bm{0}_{N \times 1} \\ \bm{0}_{1 \times N} & 0
        \end{bmatrix}, 
        \bm{F}_0 = 
        \begin{bmatrix}
            \bm{F} \\ \bm{0}_{1 \times K} 
        \end{bmatrix}. 
        \label{equ:definition of E0 Ei G0 F0}
    \end{aligned}
\end{equation}
\else
\begin{equation}
    \bm{E}_0 = 
    \begin{bmatrix}
        \bm{0}_{N \times N} & \bm{1}_{N \times 1} \\ \bm{1}_{1 \times N} & 0
    \end{bmatrix}, 
    \bm{E}_{i, i} = \diag(\bm{e}_i), 
    \bm{G}_0 = 
    \begin{bmatrix}
        \bm{G}\bm{G}\H & \bm{0}_{N \times 1} \\ \bm{0}_{1 \times N} & 0
    \end{bmatrix}, 
    \bm{F}_0 = 
    \begin{bmatrix}
        \bm{F} \\ \bm{0}_{1 \times K} 
    \end{bmatrix}. 
    \label{equ:definition of E0 Ei G0 F0}
\end{equation}
\fi
The equivalence between $\mathcal{P}_2'$ and $\mathcal{P}_3$ is proved in \textbf{Appendix \ref{Proof of the equivalence between 2 problems}}. 

After the conversion of the optimization variables, the objective function \eqref{pro:3-a} and the constraints \eqref{pro:3-b}, \eqref{pro:3-c}, and \eqref{pro:3-d} are all convex. 
Therefore, the only challenge is the non-convexity of \eqref{pro:3-e}.
If we relax the constraint \eqref{pro:3-e}, which is called SDR~\cite{Boyd2004convex,On2007So}, the optimization problem will be expressed as a convex SDP problem 
\ifx\onecol\undefined
\begin{subequations}
    \begin{align}
    \mathcal{P}_3': \min_{\bm{X}} & -\frac{1}{4}P_{\text{RIS}} \tr \left( \bm{E}_0 \bm{X} \right) + \tr \left(\bm{F}_0\H \left( \bm{X} \odot \bm{G}_0 \right) \bm{F}_0 \bm{P}^{-1}\right)^{-1}, \\ 
    \text{s.t.} \ & \tr\left( \left(\bm{F}_0\H \left( \bm{X} \odot \bm{G}_0 \right) \bm{F}_0 \bm{P}^{-1}\right)^{-1} \right) \leq \Pmax, \\ 
    & \tr \left( \bm{E}_{i, i} \bm{X} \right) = 1, \ i \in {1, 2, ..., N+1}, \\ 
    & \bm{X} \succeq 0. 
    \end{align}
\end{subequations}
\else
\begin{subequations}
    \begin{align}
    \mathcal{P}_3': \min_{\bm{X}} \ & -\frac{1}{4}P_{\text{RIS}} \tr \left( \bm{E}_0 \bm{X} \right) + \tr\left( \left(\bm{F}_0\H \left( \bm{X} \odot \bm{G}_0 \right) \bm{F}_0 \bm{P}^{-1}\right)^{-1} \right), \\ 
    \text{s.t.} \ & \tr\left( \left(\bm{F}_0\H \left( \bm{X} \odot \bm{G}_0 \right) \bm{F}_0 \bm{P}^{-1}\right)^{-1} \right) \leq \Pmax, \\ 
    & \tr \left( \bm{E}_{i, i} \bm{X} \right) = 1, \ i \in {1, 2, ..., N+1}, \\ 
    & \bm{X} \succeq 0. 
    \end{align}
\end{subequations}
\fi
The problem $\mathcal{P}_3'$ is a standard convex semi-definite programming, which can be solved with general procedures like the interior-point method by CVX solvers~\cite{cvx}. 

The next step is to acquire the low-rank solution $\bm{X}$ of $\mathcal{P}_3$ (i.e. $\bm{q}$ of $\mathcal{P}_2'$) from the solution $\tilde{\bm{X}}$ of $\mathcal{P}_3'$. 
According to the positive definiteness of $\tilde{\bm{X}}$, we can define $\bm{V} = [\bm{v}_1, \bm{v}_2, ..., \bm{v}_N] \in \mathbb{R}^{N \times N}$ as $ \bm{V}\T\bm{V} = \tilde{\bm{X}}(1:N, 1:N) $. 
Randomly select $\bm{u} \in \mathbb{R}^N$ from uniform distribution on $N$-dimensional sphere (an implementation is to construct $\tilde{\bm{u}}$ where $\tilde{u}_i \sim \mathcal{N}(0, 1)$, then $\bm{u} = \tilde{\bm{u}}/\Vert \tilde{\bm{u}} \Vert$). 
Then, the estimation $\tilde{q}_i$ is the sign of projection of $\bm{v}_i$ to $\bm{u}$, i.e. $\tilde{q}_i = \text{sgn}\left(\bm{u}\T \bm{v}_i\right)$. 
We can repeat the procedure for times to select the optimal solution. 
The method based on SDP relaxation can be summarized in \textbf{Algorithm \ref{alg:SDP relaxation}}. 
\begin{algorithm}
    \caption{SDP Relaxation}
    \label{alg:SDP relaxation}
    \setstretch{1.35}
    \begin{algorithmic}[1]
        \REQUIRE Number of RIS elements $N$; Channel matrix $\bm{F}, \bm{G}$; Energy consumption of each RIS element $P_{\text{RIS}}$. 
        \ENSURE RIS beamforming state $\bm{q}$. 

        \STATE Calculate $\bm{F}_0$, $\bm{G}_0$. 
        \STATE Solve the problem $\mathcal{P}_3'$ with the standard SDP procedures and acquire the optimal $\tilde{\bm{X}}$. 
        \STATE $\bm{V} \leftarrow \left(\tilde{\bm{X}}\left(1:N, 1:N\right)\right)^{\frac{1}{2}}$
        \FOR {$i = 1, 2, ..., N_{\text{SDP}}$} 
            \STATE Randomly choose $\tilde{\bm{u}}^{(i)}$ with $\tilde{u}_n^{(i)} \sim \mathcal{N}(0, 1)$. 
            \STATE $\bm{u}^{(i)} \leftarrow \tilde{\bm{u}}^{(i)} / \left\Vert \tilde{\bm{u}}^{(i)} \right\Vert$ 
            \STATE $\hat{q}_n^{(i)} \leftarrow \text{sgn}\left( \bm{v}_n\T \bm{u}^{(i)} \right)$ 
            \STATE $g^{(i)} \leftarrow -\frac{1}{2} P_0 \bm{1}\T \bm{q}^{(i)} + \sum_{k=1}^K p_k t_k^{(i)}$ 
        \ENDFOR
        \STATE Select the $\bm{q}^{(i)}$ as $\bm{q}$ corresponding to the minimum $g^{(i)}$. 
        
        \RETURN $\bm{q}$ 
    \end{algorithmic}
\end{algorithm}

\subsection{Complexity \& Convergence Analysis}
\label{Complexity & Convergence Analysis 2}
The original RIS analog beamforming problem $\mathcal{P}_2$ or $\mathcal{P}_2'$ is NP-hard, which means the computational complexity for the optimal solution is at least $\mathcal{O}\left( 2^N \right)$. 
This is an unacceptable thing, especially for the RIS with a large number of reflecting elements. 
Here, the computational complexity of the two methods proposed in Subsection \ref{Search with the Maximum Gradient} and \ref{SDP Relaxation} will be discussed, and it will show that these methods are computationally acceptable. 

Firstly, the complexity of \textbf{Algorithm \ref{alg:search with gradient}} is derived as follows. 
The complexity for computing \eqref{equ:gradient of g} is $\mathcal{O}\left(K^3\left(N+K\right)\right)$, which is approximate to $\mathcal{O}\left(NK^3\right)$ due to $K \ll N$ in general. 
The complexity for the sorting procedure is $\mathcal{O}\left( N \log N \right)$, which can be ignored. 
The complexity for determining the feasibility of the solution and the value of \eqref{pro:2'-a} is $\mathcal{O}\left( NKM + K^2M + K^3 \right) \simeq \mathcal{O}\left(NKM\right)$. 
Thus, assuming for $I_{\text iter}$ iterations, the computational complexity of \textbf{Algorithm \ref{alg:search with gradient}} is $\mathcal{O}\left(I_{\rm iter} \left( NK^3 + N^2KM \right)\right) \simeq \mathcal{O}\left(I_{\rm iter}N^2KM\right)$. 

Next, we will consider \textbf{Algorithm \ref{alg:SDP relaxation}}. 
The worst-case complexity for solving the SDP optimization $\mathcal{P}_3'$ with interior-point algorithm is $\mathcal{O}(( (N + 1)^2 + 1 )^{4.5}) \simeq \mathcal{O}(N^9)$\cite{Luo2010semidefinite}. 
Compared with this, all others can be ignored. 
Thus, the computational complexity of \textbf{Algorithm \ref{alg:SDP relaxation}} is $\mathcal{O}(N^9)$. 

From the discussion above, both methods are polynomial computational complexity, compared with the exponential computational complexity of the optimal solution. 
Furthermore, the computational complexity of \textbf{Algorithm \ref{alg:SDP relaxation}} is much higher than that of \textbf{Algorithm \ref{alg:search with gradient}}. 

Now, the convergence will be analyzed. 
\textbf{Algorithm \ref{alg:SDP relaxation}} contains no iteration step, which means the convergence does not need to be discussed here. 
According to \textbf{Algorithm \ref{alg:search with gradient}}, we define $\bm{q}^{(i, j)}$ as the value of $\bm{q}$ during the $i$-th iteration and before the change of $\bm{q}^{(i)}_{d_j}$. 
Due to $g(\bm{q}^{(i, j)}) \geq g(\bm{q}^{(1, j-1)})$ which is discussed in Subsection \ref{Search with the Maximum Gradient}, this method is proved converge. 

\section{Simulation Results}
\label{Simulation Results}

\subsection{Simulation Model}
\label{Simulation Model}
\begin{table*}[t]
    \caption{Simulation Parameters}
    \label{tab:simulation parameters}
    \centering
    \tabcolsep=1cm
    \begin{tabular}{|l|c|}
        \hline 
        RIS elements $N = N_1 \times N_2$ & $64 = 8 \times 8$ \\ 
        \hline 
        Static power consumption $P_{\text{static}}$ & $10\,{\rm W}$ \\ 
        \hline 
        Power consumption of each ON-state RIS element $P_0$ & $10\,{\rm mW}$ \\ 
        \hline 
        Distance between BS and RIS $d_{\rm BS}$ & $200\,{\rm m}$ \\ 
        \hline 
        Distance between users and RIS $d_{\rm UE}$ & $200\,{\rm m}$ \\ 
        \hline
        BS antennas $M_{\rm BS}$ & 8 \\
        \hline
        Number of users $K$ & 4 \\
        \hline 
        The minimum SE requirement $\SE_{\rm min}$ & $10^{-4}\,{\rm bps/Hz}$ \\
        \hline 
        Thermal noise level $n_0$ & $-174\,{\rm dBm/Hz}$ \\ 
        \hline
        Carrier frequency $f_c$ & $3.5 \,{\rm GHz}$ \\
        \hline
        Subcarrier spacing $\rm BW$ & $180 \,{\rm kHz}$ \\
        \hline
        Noise power at receiver $\sigma^2$ & ${\rm BW} \times n_0$ \\
        \hline 
    \end{tabular}
\end{table*}

In this section, simulation results will be provided to verify the effect of the proposed algorithm for maximizing energy efficiency in an MU-MISO communication scenario. 
In our simulation, the signal models and channel models are consistent with the discussion in Section~\ref{System Model}. 
In Subsection \ref{Convergence Performance}, the convergence of the algorithm will be shown with different BS transmit power. 
Then, the performance of the proposed algorithm are tested under a wide range of BS transmit power, which is shown in Section~\ref{EE with Different BS Transmit Power}. 
Finally, the impact of the number of RIS elements on EE will be shown in Subsection \ref{Impact of the Number of RIS Elements}. 

In this subsection, we will introduce the channel model and simulation parameters. 
We assume a Rician fading channel that consists of both line-of-sight (LoS) and non-line-of-sight (NLoS) components. Specifically, the channel matrix $\bm{F} (\bm{G})$ can be expressed as follows: 
\begin{equation}
    \bm{F} = z \left(\sqrt{\frac{\kappa}{\kappa + 1}}\bm{F}_{\rm LoS} + \sqrt{\frac{1}{\kappa + 1}}\bm{F}_{\rm NLoS}\right), 
\end{equation}
where $\kappa$ is the Rician factor, $z$ is the path loss related to the distance, $\bm{F}_{\rm LoS}$ represents the LoS component of the channel $\bm{F}$, and $\bm{F}_{\rm NLoS}$ represents the NLoS component of $\bm{F}$, which follows the distribution $[\bm{F}_{\rm NLos}]_{k\ell} \sim$ i.i.d. $\mathcal{CN}\left( 0, 1 \right)$. 
The LoS component $\bm{F}_{\rm LoS}$ admits an SV channel structure, i.e., 
\begin{equation}
    \bm{F}_{\rm LoS} = \sqrt{NM}\bm{a}_N \left( \theta, \varphi \right) \bm{a}_M\H \left( \theta', \varphi' \right), 
\end{equation}
where $\left( \theta, \varphi \right), \left( \theta', \varphi' \right)$ are the azimuth and elevation angles of the beam in the coordinate system of the RIS and the BS, respectively. 
The steering vectors $\bm{a}_N, N = N_1 \times N_2$ and $\bm{a}_M, M = M_1 \times M_2 $ are defined as 
\ifx\onecol\undefined
\begin{equation}
    \begin{aligned}
    \bm{a}_{\bm N}(\theta, \varphi) = & \frac{1}{\sqrt{N}} \left[1, e^{\j\pi \sin{\theta} \sin{\varphi} / \lambda}, \cdots, e^{\j\pi (N_1 - 1)  \sin{\theta} \sin{\varphi} / \lambda}\right]_{N_1}\T \\ 
    & \otimes \left[1, e^{\j \pi  \cos{\varphi} / \lambda}, \cdots, e^{\j \pi (N_2 - 1)  \cos{\varphi} / \lambda}\right]_{N_2}\T,\\
    \bm{a}_{\bm M}(\theta, \varphi) = & \frac{1}{\sqrt{M}} \left[1, e^{\j\pi  \sin{\theta} \sin{\varphi} / \lambda}, \cdots, e^{\j\pi (M_1 - 1)  \sin{\theta} \sin{\varphi} / \lambda}\right]_{M_1}\T \\ 
    & \otimes \left[1, e^{\j \pi \cos{\varphi} / \lambda}, \cdots, e^{\j \pi (M_2 - 1) \cos{\varphi} / \lambda}\right]_{M_2}\T.
    \end{aligned}
\end{equation}
\else
\begin{equation}
    \begin{aligned}
    &\bm{a}_{\bm N}(\theta, \varphi) = & \frac{1}{\sqrt{N}} \left[1, e^{\j\pi \sin{\theta} \sin{\varphi} / \lambda}, \cdots, e^{\j\pi (N_1 - 1)  \sin{\theta} \sin{\varphi} / \lambda}\right]_{N_1}\T \otimes \left[1, e^{\j \pi  \cos{\varphi} / \lambda}, \cdots, e^{\j \pi (N_2 - 1)  \cos{\varphi} / \lambda}\right]_{N_2}\T,\\
    &\bm{a}_{\bm M}(\theta, \varphi) = & \frac{1}{\sqrt{M}} \left[1, e^{\j\pi  \sin{\theta} \sin{\varphi} / \lambda}, \cdots, e^{\j\pi (M_1 - 1)  \sin{\theta} \sin{\varphi} / \lambda}\right]_{M_1}\T \otimes \left[1, e^{\j \pi \cos{\varphi} / \lambda}, \cdots, e^{\j \pi (M_2 - 1) \cos{\varphi} / \lambda}\right]_{M_2}\T.
    \end{aligned}
\end{equation}
\fi
Core simulation parameters are shown in \textbf{Table~\ref{tab:simulation parameters}}~\cite{wang2022reconfigurable,zhu2022sensing}.

\subsection{Convergence Performance}
\label{Convergence Performance}
\ifx\onecol\undefined
\begin{figure}[!t]
    \centering
    \myincludegraphics{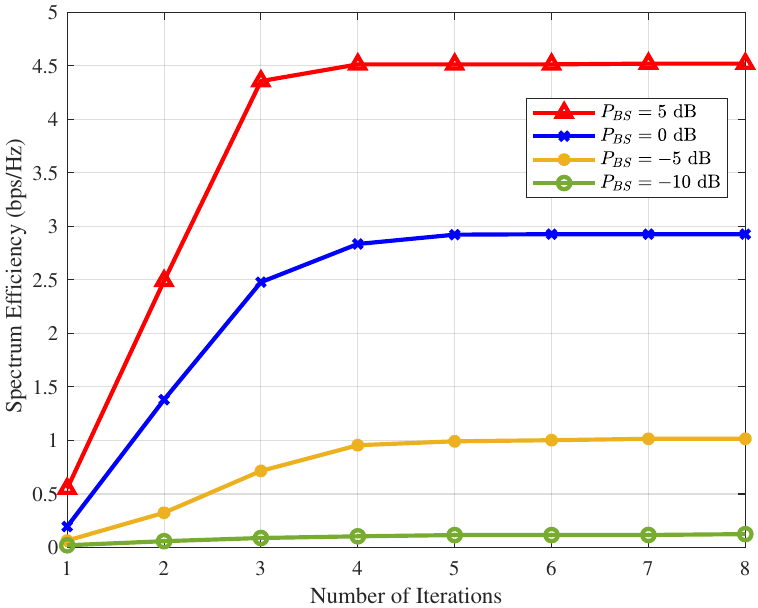}
    \caption{Convergence performances of the AO algorithm with four different BS transmit power, where the $x$-axis denotes the number of iterations, $y$-axis denotes the spectrum efficiency (bps/Hz).}
    \label{fig:Convergence Performance_1}
\end{figure}
\begin{figure}[!t]
    \centering
    \myincludegraphics{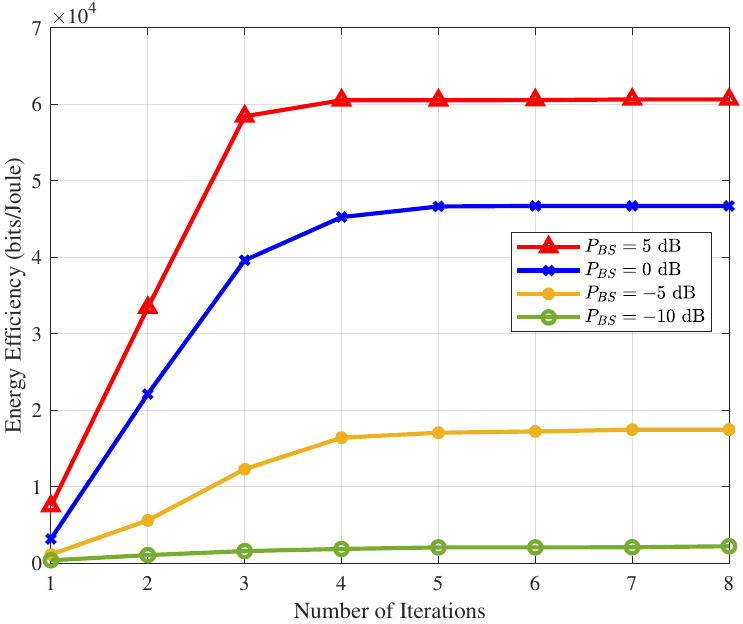}
    \caption{Convergence performances of the AO algorithm with four different BS transmit power, where the $x$-axis denotes the number of iterations, $y$-axis denotes the energy efficiency (bits/Joule).}
    \label{fig:Convergence Performance_2}
\end{figure}
\else
\begin{figure}[!t]
    \centering
    \myincludegraphics{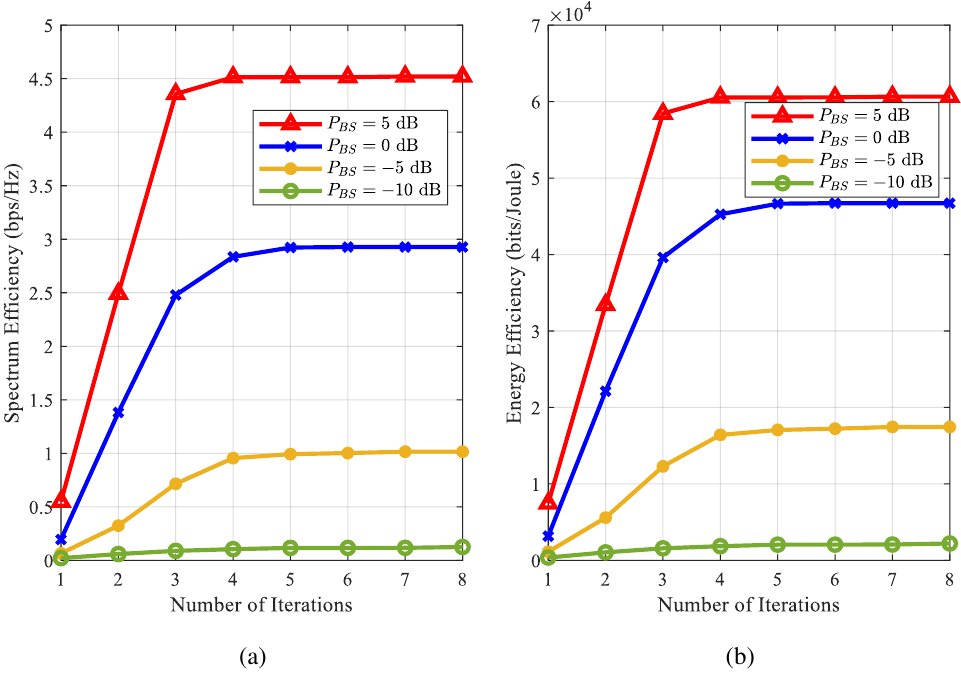}
    \caption{Convergence performances of the AO algorithm with four different BS transmit power, where the $x$-axis denotes the number of iterations. (a) Convergence of the spectrum efficiency (SE); (b) Convergence of the energy efficiency (EE).}
    \label{fig:Convergence Performance}
\end{figure}
\fi

In Section~\ref{Algorithms}, we have proposed an AO algorithm, which contains 2 steps executed iteratively: Power allocating and RIS analog beamforming. 
In Subsection~\ref{Complexity & Convergence Analysis 1}, we have theoretically proved the convergence of the algorithm. Here, we will check the convergence rate by numerical simulations. 
The simulation parameters are shown in \textbf{Table \ref{tab:simulation parameters}}. 
The spatial angular coordinates  $\left( \theta, \varphi \right)$ of BS and user (relative to the RIS) are drawn from a uniform distribution on $\left[ -\pi/2, \pi/2 \right]$ and $\left[ \pi/3, 2\pi/3 \right]$ respectively. 

The SE and EE of the system are studied as a function of the iteration number of the proposed algorithm. 
\ifx\onecol\undefined
Fig.~\ref{fig:Convergence Performance_1} shows the SE of the communication system after each iteration with different BS transmit power, and Fig.~\ref{fig:Convergence Performance_2} shows the evolution of the system EE as the iteration proceeds. 
\else
Fig.~\ref{fig:Convergence Performance} (a) shows the SE of the communication system after each iteration with different BS transmit power, and Fig.~\ref{fig:Convergence Performance} (b) shows the evolution of the system EE as the iteration proceeds. 
\fi
These figures show that after several previous iterations, the SE and EE of the system all maintain a stable value, which verifies the convergence of the AO algorithm. Simulation results have demonstrated that, the AO algorithm framework achieves convergence in only two or three iterations. Moreover, this convergence rate is almost independent of the transmitted power, thus proving that the AO algorithm framework is computationally stable and widely applicable to different communication conditions.

\subsection{EE with Different BS Transmit Power}
\label{EE with Different BS Transmit Power}

\ifx\onecol\undefined
\begin{figure}[!t]
    \centering
    \myincludegraphics{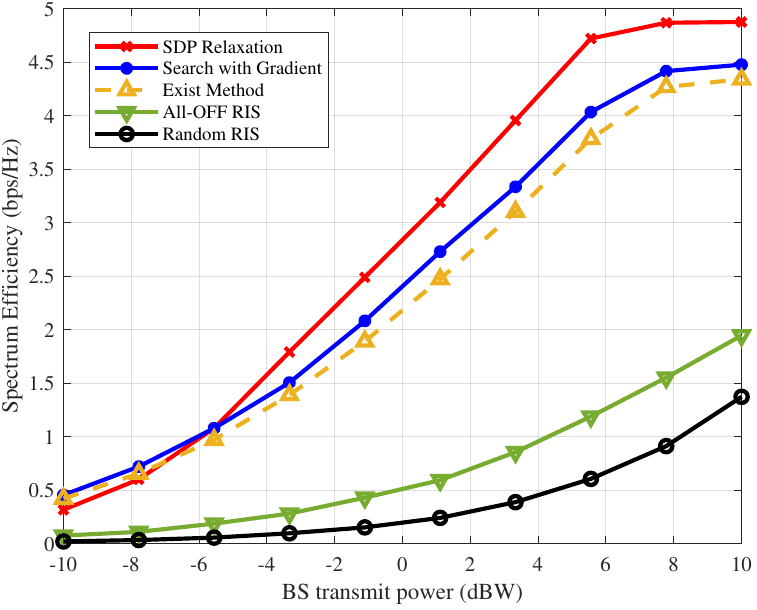}
    \caption{SE curves based on proposed algorithm, existing method~\cite{Wu2019algorithm}, and baselines; $x$-axis denotes the BS transmit power; $y$-axis denotes the spectrum efficiency (bps/Hz).}
    \label{fig:SE with BS transmit power}
\end{figure}
\begin{figure}[!t]
    \centering
    \myincludegraphics{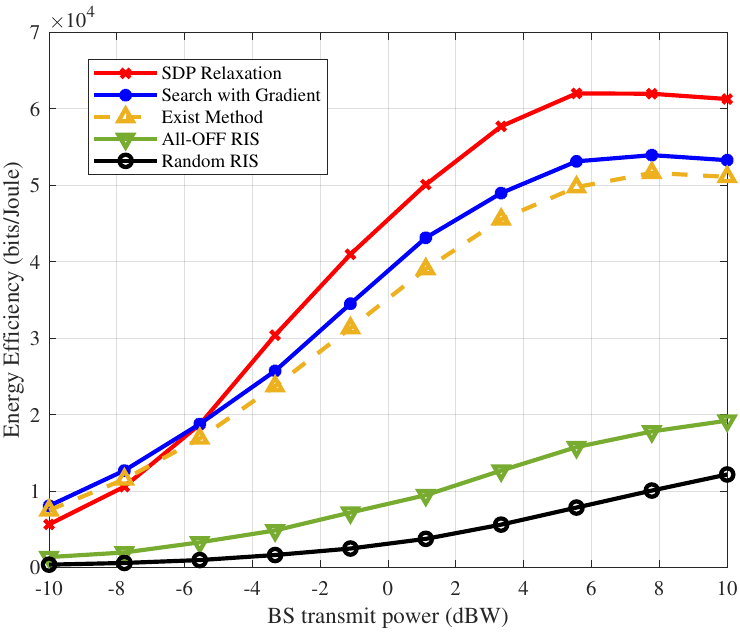}
    \caption{EE curves based on proposed algorithm, existing method~\cite{Wu2019algorithm}, and baselines; $x$-axis denotes the BS transmit power; $y$-axis denotes the energy efficiency (bits/Joule).}
    \label{fig:EE with BS transmit power}
\end{figure}
\else
\begin{figure}[!t]
    \centering
    \myincludegraphics{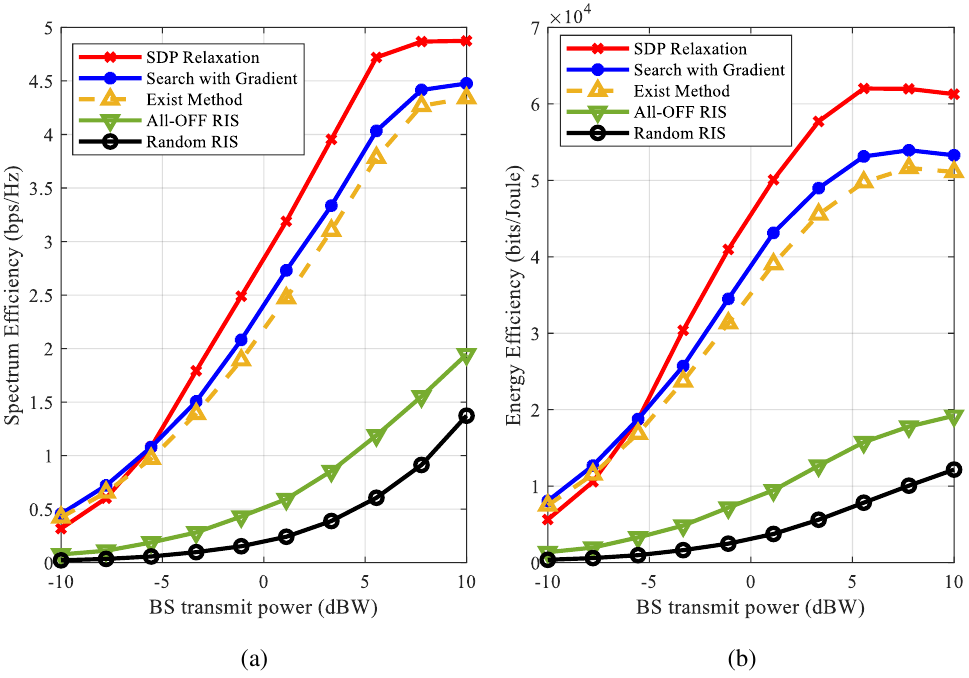}
    \caption{Performance curves based on proposed algorithm, existing method~\cite{Wu2019algorithm}, and baselines; $x$-axis denotes the BS transmit power. (a) $y$-axis denotes the spectrum efficiency (bps/Hz); (b) $y$-axis denotes the energy efficiency (bits/Joule).}
    \label{fig:SE EE with BS transmit power}
\end{figure}
\fi

In this subsection, we mainly focus on the energy efficiency optimized by the proposed algorithm as a function of the transmitted power. Note that since the thermal noise power $\sigma_n^2$ at the receiver is fixed, the transmitted power admits a constant dB difference from the transmitter \ac{SNR}. 
The simulation parameters are consistent with those in Subsection~\ref{Convergence Performance}, and the BS transmit power varies from $-10\,{\rm dBW}$ to $10\,{\rm dBW}$. 

\ifx\onecol\undefined
The spectrum efficiency and energy efficiency curves are provided in Fig.~\ref{fig:SE with BS transmit power} and Fig.~\ref{fig:EE with BS transmit power}. 
\else
As shown in Fig.~\ref{fig:SE EE with BS transmit power}, the spectrum efficiency and energy efficiency curves are provided in subgraph (a) and (b) respectively. 
\fi
The blue curves ``Search with Gradient'' in both subgraphs represent the results with the RIS analog beamforming algorithm designed in Subsection~\ref{Search with the Maximum Gradient}, and the red curves ``SDP Relaxation'' represent that in Subsection~\ref{SDP Relaxation}. 
It should be noted that both the curves ``Search with Gradient'' and ``SDP Relaxation'' belong to the AO framework. Their only difference is that, the former employs the gradient search RIS analog beamforming method, while the latter utilizes the SDP relaxation-based RIS analog beamforming method.
The black curves ``Random RIS'' and the green curves ``All-OFF RIS'' are baselines, which are given by random RIS analog beamforming and the RIS beamforming as an identity matrix (i.e. all the elements of RIS are configured to the OFF states). 
We also plot the performance of an existing method proposed in \cite{Wu2019algorithm}, which solves $\mathcal{P}_2$ via successively setting the phase shifts of all elements in order from $n = 1$ to $n = N$, and update each $\theta_n$ by fixing all other $\theta_k$'s, $\forall k \neq n$. The process does not stop until convergence.

\ifx\onecol\undefined
Fig.~\ref{fig:SE with BS transmit power} and Fig.~\ref{fig:EE with BS transmit power} show that both two methods have better SE and EE performances in typical scenarios than the two baselines, as well as the existing method. 
\else
Fig.~\ref{fig:SE EE with BS transmit power} shows that both two methods have better SE and EE performances in typical scenarios than the two baselines, as well as the existing method. 
\fi
The SDP relaxation method may achieve less performance when the \ac{BS} transmit power is extremely low, but such scenarios are rare in real-world settings. 
The reason why ``Search with Gradient" outperforms the existing method is that additional information provided by the gradient is fully utilized, thus more decrease in $\mathcal{P}_2$'s objective function can be achieved in each iteration.
There is a noteworthy phenomenon that both the SE and EE curves will reach a platform, where the value will not continuously grow as the increasing BS transmit power. 
This is because in order to maximize energy efficiency, the power BS consumed to transmit messages is expected low. 
Thus, even if the potential transmitted power is relatively high, the optimization algorithm tends to transmit with insufficient power, which causes the SE and EE curves flat when BS transmit power restriction is high enough. 

We can learn from the figure that the SDP relaxation method proposed in Subsection \ref{SDP Relaxation} is generally better than the gradient searching method proposed in Subsection~\ref{Search with the Maximum Gradient}. 
This is reasonable because the solution of optimization problem $\mathcal{P}_2'$ is provided the approximate value in the real sense only by the latter one, while the former one only provides an effective method for a better solution. 
\ifx\onecol\undefined
However, in Fig.~\ref{fig:SE with BS transmit power}, we notice that the results seem to be less significant when BS transmit power is relatively high, which is due to the balance between the increase of SE and the decrease of BS power consumption. 
\else
However, in Fig.~\ref{fig:SE EE with BS transmit power} (a), we notice that the results seem to be less significant when BS transmit power is relatively high, which is due to the balance between the increase of SE and the decrease of BS power consumption. 
\fi
In order to get a higher EE, the lower SE carrying by the lower BS power consumption is hard to avoid, and this is particularly prominent under high BS power consumption.

\subsection{Impact of the Number of RIS Elements}
\label{Impact of the Number of RIS Elements}

\ifx\onecol\undefined
\begin{figure}[!t]
    \centering
    \myincludegraphics{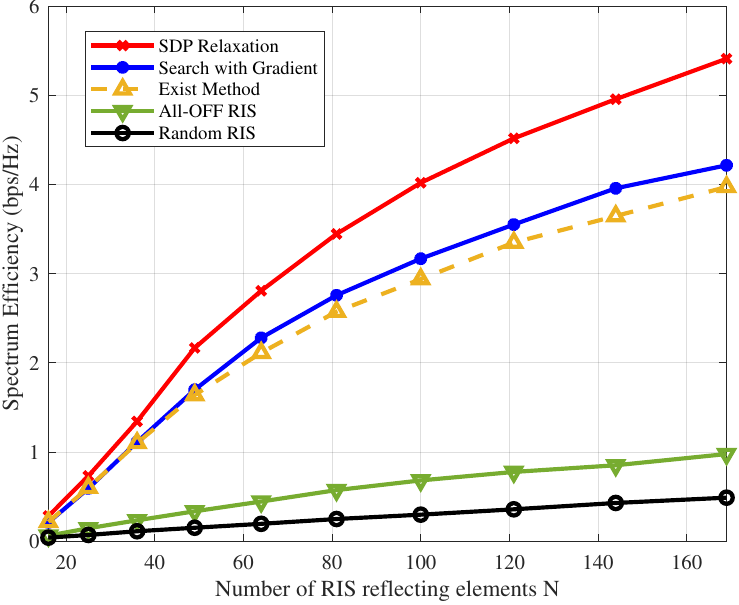}
    \caption{SE curves based on proposed algorithm, existing method~\cite{Wu2019algorithm}, and baselines; $x$-axis denotes the number of RIS reflecting elements; $y$-axis denotes the spectrum efficiency (bps/Hz).}
    \label{fig:SE with RIS elements}
\end{figure}
\begin{figure}[!t]
    \centering
    \myincludegraphics{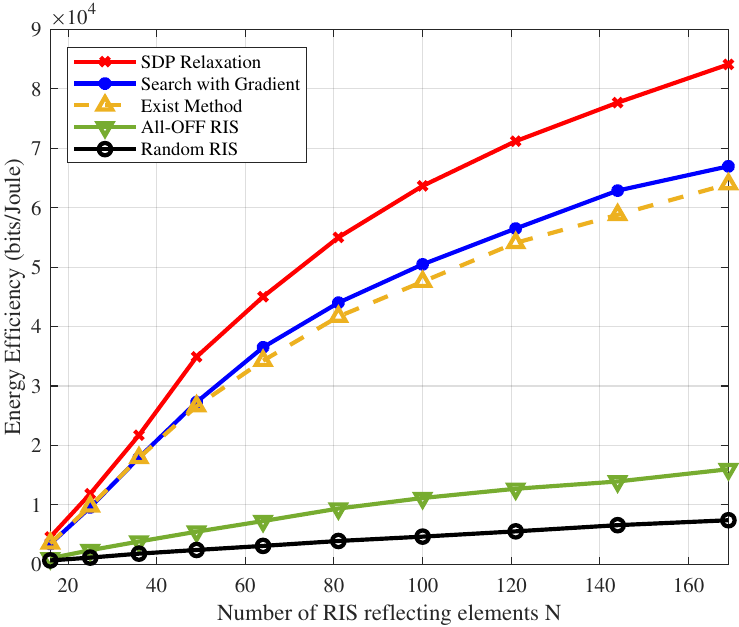}
    \caption{EE curves based on proposed algorithm, existing method~\cite{Wu2019algorithm}, and baselines; $x$-axis denotes the number of RIS reflecting elements; $y$-axis denotes the energy efficiency (bits/Joule).}
    \label{fig:EE with RIS elements}
\end{figure}
\else
\begin{figure}[!t]
    \centering
    \myincludegraphics{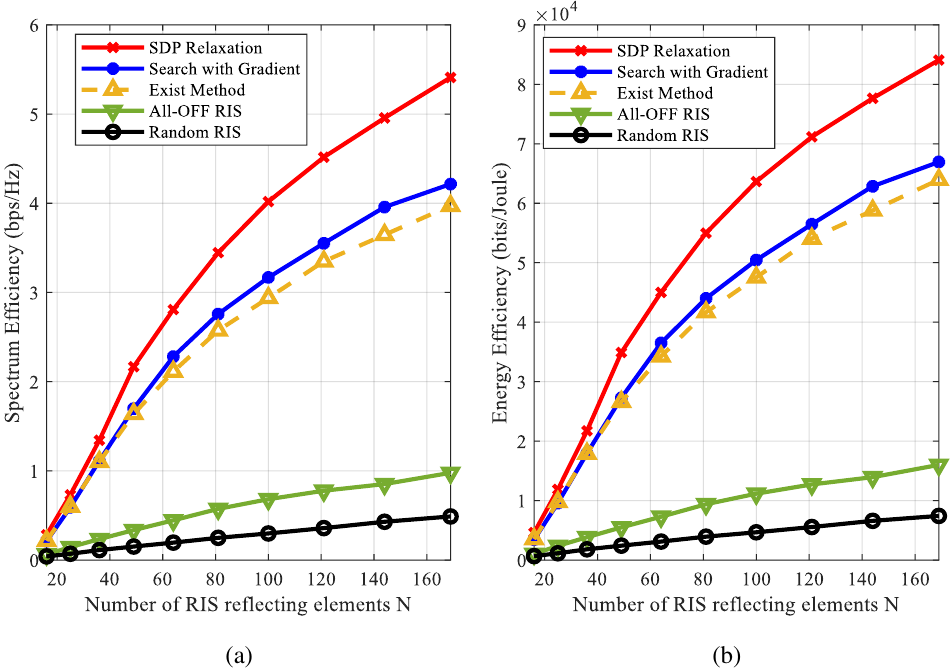}
    \caption{Performance curves based on proposed algorithm, existing method~\cite{Wu2019algorithm}, and baselines; $x$-axis denotes the number of RIS reflecting elements. (a) $y$-axis denotes the spectrum efficiency (bps/Hz); (b) $y$-axis denotes the energy efficiency (bits/Joule).}
    \label{fig:SE EE with RIS elements}
\end{figure}
\fi

Here, we mainly show the impact of the number of RIS reflecting elements on the SE and EE. 
The simulation parameters are consistent with those in Subsection \ref{Convergence Performance}, and the number of RIS reflecting elements varies from $4 \times 4$ to $13 \times 13$. 

\ifx\onecol\undefined
Fig.~\ref{fig:SE with RIS elements} and Fig.~\ref{fig:EE with RIS elements} show the SE and EE curves of the communication system by both two methods. 
\else
Fig.~\ref{fig:SE EE with RIS elements} shows the SE and EE curves of the communication system by both two methods. 
\fi
The trends of the five curves are similar to those in Subsection \ref{EE with Different BS Transmit Power}, which represents the performance of algorithms and the baselines respectively. 
\ifx\onecol\undefined
From Fig.~\ref{fig:SE with RIS elements} and Fig.~\ref{fig:EE with RIS elements}, it is obvious that the SE and EE of the communication system arise as the number of RIS reflecting elements increases, for the stronger beamforming effect by RIS. 
\else
From Fig.~\ref{fig:SE EE with RIS elements}, it is obvious that the SE and EE of the communication system arise as the number of RIS reflecting elements increases, for the stronger beamforming effect by RIS. 
\fi
Also, it shows that the optimization effect of the SDP relaxation method is better than that of the gradient searching method, and both are better than that of baselines, which accords with the above results. 
This also shows the generality of the proposed algorithm.

\section{Conclusion}
\label{Conclusion}
In this paper, in order to improve the energy efficiency of RIS-assisted MU-MISO systems, we adopted a more accurate power dissipation model for 1-bit RIS elements based on the prevailing PIN-based RIS fabrication technologies.  
Based on this improved RIS power model, the energy efficiency has been formulated and optimized by jointly minimizing the active transmitted power and the passive RIS power dissipation. 
The joint optimization problem is first formulated as a non-convex integer programming problem, and then the non-convexity is addressed by rank-1 SDP relaxation. 
Convergence of the proposed algorithm has been verified by both theoretical analysis and numerical computations.  
Simulation results have demonstrated that, the energy efficiency of the proposed joint passive-active power minimizing algorithm outperforms that of the existing methods. 

For future works, the 1-bit RIS power dissipation model can be generalized to more practical multi-bit versions for improved accuracy. The corresponding joint optimization algorithms can be re-designed to adapt to this enhanced RIS manipulation capability.

\appendices
\section{Analytical solution of the problem \eqref{equ:Dinkelbach1}}
\label{Analytical solution of the problem Dinkelbach1}
Here, we will analytically solve the optimization problem \eqref{equ:Dinkelbach1}, which is 
\begin{subequations}
    \begin{align}
        \mathcal{P}_A: \bm{P} = \arg \max_{\bm{P}} \ & \sum_{k=1}^K \log_2 \left( 1 + \frac{p_k}{\sigma_n^2} \right) - \lambda \left( P_1 + \sum_{k=1}^K p_k t_k \right), \label{pro:A-a} \\ 
        \text{s.t.} \ & \sum_{k=1}^K p_k t_k \leq \Pmax, \label{pro:A-b} \\ 
        & p_k \geq \Pmin, \ \forall k \in \mathcal{K} \label{pro:A-c}.  
    \end{align}
\end{subequations}
As shown above, the target is to maximize a concave function \eqref{pro:A-a} with affine constraints \eqref{pro:A-b} and \eqref{pro:A-c}. 
Firstly, the Lagrange function of $\mathcal{P}_A$ can be expressed as 
\begin{equation}
    \begin{aligned}    
    \mathcal{L} \left( p_k, \bm{\mu} \right) =\ & \lambda \left( P_1 + \sum_{k=1}^K p_k t_k \right) - \sum_{k=1}^K \log_2 \left( 1 + \frac{p_k}{\sigma_n^2} \right) \\
    &+ \mu_0 \left( \sum_{k=1}^K p_k t_k - \Pmax \right) - \sum_{k=1}^K \mu_k \left( p_k - p_{\text{min}} \right). 
    \end{aligned}
\end{equation}
Then, the KKT condition of $\mathcal{P}_A$ can be written as 
\begin{subequations}
    \begin{align}
    & \left( \lambda + \mu_0 \right) t_k - \mu_k - \frac{1}{\left( p_k + \sigma_n^2 \right)\log 2} = 0, \ \forall k \in \mathcal{K}, \label{equ:kkt a} \\ 
    & \mu_0 \left( \sum_{k=1}^K p_k t_k - \Pmax \right) = 0, \label{equ:kkt b} \\
    & \mu_k \left( p_k - p_{\text{min}} \right) = 0, \ \forall k \in \mathcal{K}, \label{equ:kkt c} \\ 
    & \mu_0, \mu_k \geq 0, \ \forall k \in \mathcal{K}. \label{equ:kkt d} 
    \end{align}
\end{subequations} 
These are similar to water-filling solutions in form, but the relaxation term with $\lambda$ makes some difference. 
Here, we substitute \eqref{equ:kkt c} into \eqref{equ:kkt a}, and then we obtain 
\begin{equation}
    p_k = \max \left\{ \frac{1}{\log 2} \frac{1}{\left( \lambda + \mu_0 \right) t_k} - \sigma_n^2, \Pmin \right\}. 
\end{equation}
With the equation \eqref{equ:kkt b}, we have 
\begin{equation}
    \mu_0 \left( \sum_{k=1}^K \max \left\{ \frac{1}{\log 2} \frac{1}{\lambda + \mu_0} - t_k \sigma_n^2, \Pmin t_k \right\} - \Pmax \right) = 0. 
\end{equation}
Then, after introducing two auxiliary variables, the analytical solution of $\mathcal{P}_A$ can be written as 
\begin{subequations}
    \begin{align}
    & \zeta: \sum_{k=1}^K \max\left\{ \zeta - t_k \sigma_n^2, t_k p_{\text{min}} \right\} = \Pmax, \\
    & \xi = \min \left\{ \zeta, \frac{1}{\lambda \log 2} \right\}, \\
    & p_k = \max \left\{ \frac{1}{t_k} \left( \xi - t_k \sigma_n^2 \right), p_{\text{min}} \right\}. 
    \end{align}
\end{subequations}

\section{Proof of the equivalence between $\mathcal{P}_2'$ and $\mathcal{P}_3$}
\label{Proof of the equivalence between 2 problems}
In order to simplify the problem, we define the variable matrix $\bm{X}$ in order to replace $\bm{q}$, which can be expressed as 
\begin{equation}
    \bm{X} = 
    \begin{bmatrix}
        \bm{q} \\ 
        1
    \end{bmatrix}
    \begin{bmatrix}
        \bm{q}\T & 1
    \end{bmatrix} = 
    \begin{bmatrix}
        \bm{q}\bm{q}\T & \bm{q} \\ 
        \bm{q}\T & 1 
    \end{bmatrix}, 
    \label{equ:definition of X}
\end{equation}
where $\bm{q}\bm{q}\T$ can replace the quadratic term, and $\bm{q}$ or $\bm{q}\T$ can replace the linear term. 
The expression of $\bm{H}\H\bm{H}$ can be derived as follows  
\begin{equation}
    \begin{aligned}
    \bm{H}\H \bm{H} & = \bm{F}\H \bm{\Theta} \bm{G G}\H \bm{\Theta F} \\ 
    & = \bm{F}\H \diag \left( \bm{q} \right) \bm{G G}\H \diag \left( \bm{q} \right) \bm{F}. 
    \end{aligned}
\end{equation}
Then, we consider the $(i, j)$-element of matrix $\diag \left( \bm{q} \right) \bm{G G}\H \diag \left( \bm{q} \right)$, which is 
\ifx\onecol\undefined
\begin{equation}
    \begin{aligned}
        & \left[ \diag \left( \bm{q} \right) \bm{G G}\H \diag \left( \bm{q} \right) \right]_{(i, j)} \\
        & = \sum_{k=1}^N \sum_{l=1}^N \left[ \diag \left( \bm{q} \right) \right]_{(i, k)} \left[ \bm{G}\bm{G}\H \right]_{(k, l)} \left[ \diag \left( \bm{q} \right) \right]_{(l, j)} \\ 
        & = \sum_{k=1}^N \sum_{l=1}^N q_i q_j \left[ \bm{G}\bm{G}\H \right]_{(k, l)} \delta_{ik} \delta_{lj} \\ 
        & = q_i q_j \left[ \bm{G}\bm{G}\H \right]_{(i, j)}, 
    \end{aligned}
\end{equation}
\else
\begin{equation}
    \begin{aligned}
        \left[ \diag \left( \bm{q} \right) \bm{G G}\H \diag \left( \bm{q} \right) \right]_{(i, j)} & = \sum_{k=1}^N \sum_{l=1}^N \left[ \diag \left( \bm{q} \right) \right]_{(i, k)} \left[ \bm{G}\bm{G}\H \right]_{(k, l)} \left[ \diag \left( \bm{q} \right) \right]_{(l, j)} \\ 
        & = \sum_{k=1}^N \sum_{l=1}^N q_i q_j \left[ \bm{G}\bm{G}\H \right]_{(k, l)} \delta_{ik} \delta_{lj} \\ 
        & = q_i q_j \left[ \bm{G}\bm{G}\H \right]_{(i, j)}, 
    \end{aligned}
\end{equation}
\fi
so the matrix above can be expressed as the Hadamard product $\bm{qq}\T \odot \bm{GG}\H$, which can be written as a linear combination of $\bm{X}$. 
Then, the matrix $\bm{H}\H\bm{H}$ can be expressed as 
\begin{equation}
    \bm{H}\H \bm{H} = \bm{F}_0\H \left( \bm{X} \odot \bm{G}_0 \right) \bm{F}_0, 
\end{equation}
where the $\bm{F}_0$ and $\bm{G}_0$ are defined in \eqref{equ:definition of E0 Ei G0 F0}. 
Then, the constraint \eqref{pro:2'-b} can be written as 
\ifx\onecol\undefined
\begin{equation}
    \begin{aligned}
        & \tr \left( \bm{P}^{\frac{1}{2}} \left( \bm{H}\H \bm{H} \right)^{-1} \bm{P}^{\frac{1}{2}} \right) \\ 
        & = \tr\left( \left(\bm{F}_0\H \left( \bm{X} \odot \bm{G}_0 \right) \bm{F}_0 \bm{P}^{-1}\right)^{-1} \right) \leq \Pmax. 
    \end{aligned}
\end{equation}
\else
\begin{equation}
    \tr \left( \bm{P}^{\frac{1}{2}} \left( \bm{H}\H \bm{H} \right)^{-1} \bm{P}^{\frac{1}{2}} \right) = \tr\left( \left(\bm{F}_0\H \left( \bm{X} \odot \bm{G}_0 \right) \bm{F}_0 \bm{P}^{-1}\right)^{-1} \right) \leq \Pmax. 
\end{equation}
\fi

The other constraint \eqref{pro:2'-c} of $\mathcal{P}_2'$ can also be written as 
\begin{equation}
    q_n^2 = 1, \ \forall n \in \mathcal{N}, 
\end{equation}
which means the diagonal of $\bm{X}$ (i.e. $q_n^2$) is always $1$, which leads to $N+1$ constraints 
\begin{equation}
    \tr \left( \bm{E}_{i, i} \bm{X} \right) = 1. 
\end{equation}
In addition, according to \eqref{equ:definition of X}, the semi-definite constraint $\bm{X} \succeq 0$ and the rank-one constraint $\text{rank}\left(\bm{X}\right) = 1$ should be considered in order to ensure the equivalence. 
This completes the proof.

\footnotesize
\balance
\bibliographystyle{IEEEtran}
\bibliography{BinaryRIS, IEEEabrv}

\end{document}